\documentclass[12pt]{article}
\pdfoutput=1
\usepackage[english]{babel}
\usepackage{amsfonts}
\usepackage{booktabs}
\usepackage[text={17.0cm,23.7cm}]{geometry}
\usepackage{soul}
\usepackage{graphicx}
\usepackage{amsmath}
\usepackage{amssymb}

\usepackage{eurosym}
\usepackage{slashed}
\usepackage{hyperref}
\usepackage{wasysym}
\usepackage{float}
\usepackage{placeins}
\usepackage{rotating}
\usepackage{tabularx}
\usepackage{cite}

\newcommand{\DM}{\mathrm{DM}}

\begin{document}

\title{\vspace{-2cm}
{\normalsize
\flushright TUM-HEP 1079/17\\ KIAS-P17060\\}
\vspace{0.6cm}
\bf Optimized velocity distributions for direct dark matter detection \\ [8mm]}

\author{Alejandro Ibarra$^{1,2}$, Andreas Rappelt$^{1}$ \\[2mm]
{\normalsize\it $^1$ Physik-Department T30d, Technische Universit\"at M\"unchen,}\\[-0.05cm]
{\it\normalsize James-Franck-Stra\ss{}e, 85748 Garching, Germany}\\[2mm]
{\normalsize\it $^2$ School of Physics, Korea Institute for Advanced Study, Seoul 02455, South Korea}
}

\date{}

\maketitle
\begin{abstract}
	We present a method to calculate, without making assumptions about the local dark matter velocity distribution, the maximal and minimal number of signal events in a direct detection experiment given a set of constraints from other direct detection experiments and/or neutrino telescopes. The method also allows to determine the velocity distribution that optimizes the signal rates. We illustrate our method with three concrete applications: {\it i)} to derive a halo-independent upper limit on the cross section from a set of null results, {\it ii)} to confront in a halo-independent way a detection claim to a set of null results and {\it iii)} to assess, in a halo-independent manner, the prospects for detection in a future experiment given a set of current null results.
\end{abstract}

\section{Introduction}
\label{sec:intro}

Numerous observations point towards the existence of a population of a non-luminous matter component in galaxies, clusters of galaxies and the Universe at large scales, dubbed dark matter (for reviews, see ~\cite{Bertone:2010zza,Bergstrom:2000pn,Bertone:2004pz}). A plausible hypothesis for the nature of the dark matter is that it is constituted by new particles not contained in the Standard Model. If correct, dark matter particles would scatter on nuclei, leading to potential tests of the particle dark matter hypothesis and to the eventual determination of its characteristics, such as its mass and the strength of its interactions with ordinary matter. 

Various search strategies have been proposed based on the possibility that dark matter particles could scatter with nuclei. Direct detection experiments aim to detect the nuclear recoil induced by the elastic scattering of the dark matter particles traversing a detector at the Earth~\cite{Goodman:1984dc}. This search strategy requires an exquisite suppression of the rate of recoils by electromagnetic interactions of $\alpha$-particles, electrons, and photons produced by the radioactive isotopes in the surrounding material, as well as by nuclear interactions of neutrons produced by natural radioactivity (for a review, see {\it e.g.}~\cite{Undagoitia:2015gya}). Alternatively, one may search for the characteristic annual modulation of the rate of dark matter induced scatterings against the, mostly time-independent, rate of background events~\cite{Drukier:1986tm,Freese:2012xd}. A complementary strategy consists in the search, using neutrino telescopes, of a flux of high energy neutrinos correlated to the direction of the Sun, which is hypothetically produced in the annihilation of dark matter particles previously captured in the Sun via a series of scatterings with the matter in the solar interior~\cite{Silk:1985ax}. 

The interpretation of the experimental results in a concrete model of particle dark matter suffers from various nuclear and astrophysical uncertainties. Concretely, the rate of dark matter-nucleus scatterings crucially depend on the flux of dark matter particles impinging the target material, which in turn depends on the dark matter number density and velocity distribution inside the Solar System. It is common in the literature to adopt a local dark matter density $\rho_{\rm loc}\approx 0.3\,{\rm GeV}/{\rm cm}^3$ and a velocity distribution in the galactic rest frame with a Maxwell-Boltzmann form. While the adopted value of the local dark matter density is well motivated by astronomical observations (see {\it e.g.} \cite{Read:2014qva}), the form of the velocity distribution is totally unknown and relies purely on theoretical considerations. The Maxwell-Boltzmann form arises as the solution to the collisionless Boltzmann equation for a dark matter distribution consisting on an isotropic, isothermal sphere with density distribution $\rho(r)\sim r^{-2}$~\cite{Drukier:1986tm}. On the other hand, N-body simulations indicate that a Maxwellian distribution might not provide a good description of the smooth halo component~\cite{Colin:1999bh,Hansen:2005yj,Vogelsberger:2008qb,Kuhlen:2009vh,Lisanti:2010qx,Mao:2012hf}. Furthermore, it has been argued that the dark matter halo of our Galaxy might contain tidal streams or a dark disk component ~\cite{Read:2008fh,Read:2009iv,Purcell:2009yp,Ling:2009eh} which may induce significant deviations in the velocity distribution from the Maxwell-Boltzmann form and in turn affect the interpretation of dark matter search experiments \cite{Green:2010gw,Peter:2011eu,Freese:2003na,Gondolo:2005hh,Savage:2006qr} (however, the existence of a dark disk in the Milky Way has been questioned in  \cite{Schaller:2016uot,Bozorgnia:2016ogo,Kelso:2016qqj} and seems to be disfavored by observations~\cite{Bovy:2013raa}). More recently,  hydrodynamical simulations have suggested that the average velocity distribution at the position of the Solar System may be well described by a Maxwell-Boltzmann form~\cite{Bozorgnia:2016ogo,Kelso:2016qqj,Sloane:2016kyi}. However, this conclusion is based on a sample of particles enclosed in a fairly large volume and deviations from the Maxwellian form at the scales of the Solar System cannot be precluded.  

Our current ignorance of the dark matter velocity distribution inside the Solar System represents an important source of uncertainty in the analysis of direct detection experiments and motivates the development of halo-independent methods. In  \cite{Fox:2010bu, Fox:2010bz} it was proposed a method to map experimental signals from one detector to another by introducing a variable that includes both the information about the dark matter scattering cross section and the integrated velocity distribution, often denoted as $\eta(v_\text{min})$. This method has been applied to quantify the compatibility of a positive claim with a null result in a halo-independent way, by comparing measurements and upper limits of this variable from different direct detection experiments~\cite{McCabe:2010zh,McCabe:2011sr,Frandsen:2011gi,Gondolo:2012rs,HerreroGarcia:2012fu,DelNobile:2013cva,Fox:2014kua,Feldstein:2014ufa,Anderson:2015xaa,Bozorgnia:2014gsa}. Other works focus on the dark matter parameter estimation in case a positive signal is detected in a direct detection experiment, either using general parametrizations of the velocity distribution \cite{Peter:2011eu,Kavanagh:2012nr,Kavanagh:2013wba,Kavanagh:2016xfi}, or by decomposing the velocity distribution into a number of streams~\cite{Feldstein:2014gza,Kahlhoefer:2016eds}. Other methods exploit the complementarity of direct search experiments and neutrino telescopes in probing different parts of the dark matter velocity distribution, and have derived halo-independent constraints on the dark matter properties from combining two null search experiments~\cite{Ferrer:2015bta}, or investigated the implications of a positive signal in a direct detection experiment for the forecast neutrino flux from the Sun ~\cite{Blennow:2015oea} or, in the event that a signal is also detected at a neutrino telescope, for the reconstruction of the dark matter parameters ~\cite{Kavanagh:2014rya}.

In this paper we develop a new method to compare, in a halo independent manner, the outcome of two or more experiments probing the dark matter distribution inside the Solar System. Utilizing the fact that the flux of dark matter particles, and therefore the rate of scatterings with nuclei, is linear in the velocity distribution, we apply techniques of linear programming to determine the velocity distribution that minimizes/maximizes the outcome of one experiment subject to the constraints from other experiments. We also illustrate this approach with three concrete applications:  {\it i)} to derive a halo-independent upper limit on the cross section from a set of null results, {\it ii)} to confront in a halo-independent way a detection claim to a set of null results and {\it iii)} to assess, in a halo-independent manner, the prospects for detection in a future experiment given a set of current null results.

This work is organized as follows. In Section \ref{sec:approaches} we review the formalism to calculate the rate and the modulation signal of dark matter induced scatterings at a direct detection experiment, as well as the dark matter capture rate in the Sun. In Section \ref{sec:optimization} we present our method to optimize the outcome of an experiment given the constraints from other experiments, and in Section \ref{sec:applications} we discuss some concrete applications of our method. Finally, in Section \ref{sec:conclusions} we present our conclusions and an outlook.

\section{Approaches to dark matter detection in the Solar System}
\label{sec:approaches}

We postulate that the dark matter distribution inside the Solar System is spatially homogeneous and has density $\rho_{\rm loc}$, is constant over time, and has a velocity distribution relative to the solar frame $f(\vec v)$ normalized such that
\begin{equation}
\int_{v~\leq v_{\text{max}}}\;dv^{3}\;f(\vec{v})=1,\label{eq:VDNormalization}
\end{equation}
where $v\equiv |\vec v|$ and $v_{\text{max}}$ is the maximal velocity of a dark matter particle that is gravitationally bound to the galaxy, expressed in the solar frame. Throughout the paper, we will adopt $v_\text{max} \simeq 777$ km/s, which is the sum of the galactic escape velocity $\simeq 533$ km/s~\cite{Piffl:2013mla} and the local velocity of the Sun with respect to the halo $\simeq 244$ km/s~\cite{Xue:2008se,McMillan:2009yr,Bovy:2009dr}. 

Three different methods have been proposed to test the particle nature of the putative dark matter population inside the solar system:  the search for dark matter induced nuclear recoils in a direct detection experiment~\cite{Goodman:1984dc}, the search for the characteristic annual modulation signal in the rate of nuclear recoils~\cite{Freese:2012xd} and the search for a high energy neutrino flux from the Sun from the annihilation of dark matter particles captured in the solar interior via scatterings~\cite{Silk:1985ax}.
 	
 The rate of nuclear recoils induced by scatterings of dark matter particles traversing a detector at the Earth can be calculated from:
\begin{align}
	R= \sum_i \int_{0}^\infty \text{d}E_R \, \epsilon_i (E_R) \frac{\xi_i \rho_\text{loc}}{m_{A_i} m_{\text{DM}}} \int_{v^{\rm (D)} \geq v_{\text{min},i}^{\rm (D)}(E_R)} \text{d}^3 v^{\rm (D)} \, v^{\rm (D)} f (\vec{v}^{\,\rm (D)}+\vec{v}_{\rm obs}(t)) \, \frac{\text{d}\sigma_i}{\text{d}E_R} \,.
\label{eq:scattering_rate}
\end{align}
Here, $\vec v^{\,\rm (D)}$ denotes the dark matter velocity in the frame of the detector D, hence the velocity distribution of dark matter particles is $f (\vec{v}^{\,\rm (D)}+\vec{v}_{\rm obs}(t))$, with $\vec{v}_{\rm obs}(t)$ the (time-dependent) velocity of the observer relative to the Sun, given by \cite{Gelmini:2000dm}:
\begin{align}
\vec{v}_{\text{obs}}=v_{\oplus}\cdot\left\{\vec{e}_{1}\cdot\sin(\omega\cdot(t-t_{\text{phase}}))-\vec{e}_{2}\cdot\cos(\omega\cdot(t-t_{\text{phase}}))\right\},\label{eq:Vobs}
\end{align}
where $v_{\oplus}=29.8~\text{km/s}$ is the absolute value of the Earth's velocity with respect to the Sun, $\omega=2\pi/\text{year}$, $t_{\text{phase}}=0.218$ is the time of Spring equinox in units of years, and $\vec{e}_{1}$ and $\vec{e}_{2}$ are unit vectors in the direction of the Sun during Spring equinox and Summer solstice, which in galactic coordinates read $\vec{e}_{1}~=~(-0.0670,0.4927,-0.8676)$,
$\vec{e}_{2}~=~(-0.9931,-0.1170,0.01032)$~\cite{Green}.
Besides, $\text{d}\sigma_i/\text{d}E_R$ is the differential cross section for the elastic scattering of a dark matter particle off a nuclear isotope $i$ with mass $m_{A_i}$ and mass fraction $\xi_i$ in the detector, producing in the scattering a nucleus with energy $E_R$. Furthermore, $v_{\text{min},i}^{\rm (D)}(E_R) = \sqrt{m_{A_i} E_R/(2 \mu_{A_i}^2)}$ is the minimal velocity necessary for a dark matter particle to induce a recoil with energy $E_R$,  with $\mu_{A_i}$ being the reduced mass of the dark matter-nucleus scattering, and $\epsilon_i(E_R)$ is the probability to detect a nuclear recoil off the target nucleus $i$ with energy $E_R$. Finally, the number of expected recoil events at a direct detection experiment reads ${\cal N}=R\cdot \mathcal{E}$, with  $\mathcal{E}$ the exposure of the experiment.

Due to the changing direction of the Earth's velocity relative to the Sun's velocity, the flux of dark matter particles at Earth is expected to change with a one-year periodicity, leading to an annual modulation in the rate of nuclear recoils which could be detected in an experiment with sufficiently long exposure~\cite{Freese:2012xd}. The modulation signal is defined as the difference of the recoil rates at June 1st and December 1st, averaged over the energy interval $[E_-,E_+]$, namely:
\begin{align}
S_{[E_-,E_+]} = \frac{1}{E_{+}-E_{-}} \cdot \frac{1}{2} \cdot \left( R_{[E_-,E_+]}\Big|_\text{June 1st} - R_{[E_-,E_+]}\Big|_ \text{Dec 1st} \right) \,,
\label{eq:modulation-DAMA}
\end{align}
where $R_{[E_-,E_+]}(t)$ is the total event rate in that bin at the time $t$, which can be calculated from Eq.~(\ref{eq:scattering_rate}). This definition is motivated by the fact that in the Standard Halo Model the maximum rate of high energetic recoils is expected around June 1st and the minimum around December 1st; in this case, $S_{[E_-,E_+]}$ can be identified with the amplitude of the annual modulation signal.

Finally, dark matter particles traversing the Sun could scatter off nuclei in the solar interior, lose energy and eventually sink to the center. This process leads to an overdensity of dark matter particles in the solar interior, where annihilations can occur at a rate which can be sufficiently large to allow observation at Earth of the high energy neutrinos produced in the annihilation ~\cite{Silk:1985ax}. Assuming that dark matter captures and annihilations occur at the same rate in the solar interior,  the neutrino flux from annihilations in the Sun is completely determined by the capture rate,  which is given by~\cite{Gould:1987ir}
\begin{align}
	C=\sum_i \int_0^{R_\odot} 4\pi r^2 \text{d}r \, \eta_i(r) \frac{\rho_\text{loc}}{m_\text{DM}} \int_{v \leq v_{\text{max},i}^{\text{(Sun)}}(r)} \text{d}^3 v \, \frac{ f (\vec{v})}{v} &\left(v^2+\left[v_\text{esc}(r)\right]^2 \right) \nonumber \times \\ 
	&\int_{m_\text{DM} v^2 /2}^{2 \mu_{A_i}^2 \left(v^2+\left[v_\text{esc}(r)\right]^2 \right)/m_{A_i}} \text{d} E_R \, \frac{\text{d} \sigma_i}{\text{d}E_R} \,,
\label{eq:general_formula_capture_rate}
\end{align}
where $\eta_i(r)$ is the number density of the element $i$ at a distance $r$ from the solar center (for which we adopt the solar model AGSS09~\cite{Serenelli:2009yc}),  $R_\odot$ is the solar radius, $v_\text{esc}(r)$ is the escape velocity from the Sun at the distance $r$ from the center, and $v_{\text{max},i}^{\text{(Sun)}}(r) = 2 \, v_\text{esc}(r) \sqrt{m_\text{DM} m_{A_i}}/\left| m_\text{DM} - m_{A_i}\right|$  is the maximum velocity of a dark matter particle such that the capture in the Sun at $r$ remains kinematically possible after scattering with the element $i$.

The theoretical interpretation of the outcome of any of the three search strategies described above is subject to uncertainties from our ignorance of the nature of the dark matter particle and its interactions with nuclei, as well as of the density and velocity distribution inside the Solar System.  It is common in the literature to cast the differential cross section as (see {\it e.g.}~\cite{Cerdeno:2010jj})
\begin{align}
\frac{\text{d}\sigma_i}{\text{d}E_R}=\frac{m_{A_i}}{2\mu_{A_i}^2v^{{\rm(D)}\,2}}(\sigma_\text{SI}F_{\text{SI},i}^2(E_R)+\sigma_\text{SD}F_{\text{SD},i}^2(E_R)) \,,
\label{eq:x-section}
\end{align}
where $\sigma_\text{SI}$ and $\sigma_\text{SD}$ are, respectively, the spin-independent (SI) and spin-dependent (SD) cross sections at zero momentum transfer, which can be calculated in a concrete dark matter model in terms of its fundamental parameters, while $F_{\text{SI},i}(E_R)$ and $F_{\text{SD},i}(E_R)$  are form factors that depend on the nucleus. In our work, we will assume that the scattering cross sections off protons and off neutrons are identical, and we will adopt the form factors reported in~\cite{Lewin:1995rx} for the SI scattering, in~\cite{Klos:2013rwa} for the SD scattering off the nuclei relevant for direct detection experiments, and in \cite{Catena:2015uha} for the SD scattering off the nuclei relevant for the dark matter capture in the Sun. Therefore,  we will treat as free parameters the dark matter mass $m_{\rm DM}$ and the SI and SD cross sections at zero momentum transfer, $\sigma_\text{SI}$ and $\sigma_\text{SD}$. Besides, it is common to assume a local dark matter density $\rho_{\rm loc}=0.3\,{\rm GeV}/{\rm cm}^3$ and a velocity distribution in the galactic rest frame following a Maxwell-Boltzmann distribution. Under these assumptions, the results from experiments can be cast as limits on the SI or SD cross sections as a function of the dark matter mass (or allowed regions, for experiments reporting a positive signal). While the Maxwell-Boltzmann form for the velocity distribution can be justified theoretically for a dark matter population in thermal equilibrium with density distribution  $\rho(r)\sim r^{-2}$, significant deviations from this simple structure cannot be precluded, especially at small scales as is the case of the Solar System. In the next section we will present a method to derive limits on the cross section from combining information of various direct detection experiments and/or neutrino telescopes, and which does not rely on the choice of the velocity distribution.

\section{Signal optimization}
\label{sec:optimization}

Our goal is to optimize the outcome of an experiment $A$, $N^{(A)}$ (where $N$ can be the recoil rate, $R$, the modulation signal, $S$, or the capture rate, $C$), given the upper limits on the outcome of $p$ experiments $B_\alpha$, $N^{(B_\alpha)}\leq N^{(B_\alpha)}_{\rm max}$ with $\alpha=1,..., p$, and the lower limits on the outcome of $q$ experiments  $B_\alpha$, $N^{(B_\alpha)}\geq N^{(B_\alpha)}_{\rm min}$ with $\alpha=p+1,..., p+q$, with the requirement that the velocity distribution is normalized to unity.

 To this end, we use the identity
\begin{equation}
f(\vec{v})=\int_{|\vec{v}_{0}|\leq v_{\text{esc}}}d^{3}v_{0}~f(\vec{v}_{0})~\delta(\vec{v}-\vec{v}_{0})\;,\label{eq:FdecompInt}
\end{equation}
which physically can be interpreted as a decomposition of the dark matter velocity distribution into a superposition of streams with fixed velocity $\vec{v}_{0}$ (and velocity distribution $f_{\vec v_0}=\delta(\vec{v}-\vec{v}_{0})$) and with weight $f(\vec v_0)$. Each stream produces in a given experiment the outcome $N_{\vec{v}_0}$, and the outcome produced by the true velocity distribution $f(\vec{v})$ can then be obtained multiplying the contribution from each stream by its weight, $f(\vec{v}_{0})$, and summing over all the streams conforming the velocity distribution. Namely,
\begin{align}
N~&=~\int_{|\vec{v}_{0}|\leq v_{\text{esc}}}\text{d}^{3}v_{0}~f(\vec{v}_{0})~N_{\vec{v}_{0}}\;.
\label{eq:StreamR}
\end{align}

The optimization problem can then be written as:
	\begin{align}
	\text{optimize}&~~{\cal F}[f] \equiv\int d^3 v_0 ~f(\vec v_0)~N^{(A)}_{\vec v_0}\;, \\
	\text{subject to}&~~\int d^3 v_0 ~f(\vec v_0)=1\;, \label{eq:norm_alt}\\
        \text{and}&~~ \int d^3 v_0 ~f(\vec v_0)~ N^{(B_\alpha)}_{\vec v_0}  \leq N^{(B_\alpha)}_{\rm max}, ~~ \alpha=1,..., p \;,\label{eq:max_dis}\\
       	\text{and}&~~\int d^3 v_0~ f(\vec v_0)~ N^{(B_\alpha)}_{\vec v_0}  \geq N^{(B_\alpha)}_{\rm min}, ~~ \alpha=p+1,..., p+q\;,\label{eq:min_alt}\\
	\text{and}&~~f(\vec v_0)\geq 0\;.\label{eq:pos_alt}
	\end{align}
where ${\cal F}[f] $ is a functional of the velocity distribution. It is important to note that the objective functional ${\cal F}[f]$ and all the constraints are linear in the velocity distribution $f$, therefore the optimization using calculus of variations does not provide a solution.~\footnote{This can be checked explicitly by deriving the Euler-Lagrange equations for the functional ${\cal F}[f] $ including  Karush-Kuhn-Tucker multipliers, in order to incorporate the inequality constraints. {\it A posteriori}, the failure of the calculus of variations to find the solution to this optimization problem can be attributed to the fact that the optimized solution turns out to be not continuously differentiable, as we will show below.} We will then apply linear programming techniques to find the velocity distribution that optimizes the signal at the experiment $A$ with the constraints listed above. To this end, we first discretize the velocity distribution into a finite sum of $n$ streams with velocity $\vec v_i$, $i=1,..., n$:
\begin{align}
f(\vec v)=\sum_{i=1}^{n} c_{\vec v_i} ~\delta(\vec{v}-\vec{v}_{i})\;,
\end{align}
with $c_i$ the weight of the stream $f_{\vec v_i}=\delta(\vec{v}-\vec{v}_{i})$ in the velocity distribution. 

Then, the discretized optimization problem can be written as:
	\begin{align}
	\text{optimize}&~~ F(c_{\vec v_1}..., c_{\vec v_n})=\sum_{i=1}^n  c_{\vec v_i} N^{(A)}_{\vec v_i}\;, \\
	\text{subject to}&~~\sum_{i=1}^n c_{\vec v_i}=1\;, \label{eq:norm_dis}\\
        \text{and}&~~ \sum_{i=1}^n c_{\vec v_i} N^{(B_\alpha)}_{\vec v_i}  \leq N^{(B_\alpha)}_{\rm max}, ~~ \alpha=1,..., p\;, \label{eq:max_dis2}\\
       	\text{and}&~~\sum_{i=1}^n c_{\vec v_i} N^{(B_\alpha)}_{\vec v_i}  \geq N^{(B_\alpha)}_{\rm min}, ~~ \alpha=p+1,..., p+q \;,\label{eq:min_dis2}\\
	\text{and}&~~c_{\vec v_i}\geq 0, ~~i=1... n\;, \label{eq:pos_dis}
	\end{align}
where $F(c_{\vec v_1}..., c_{\vec v_n})$ is a function of the weights. Finally, after straightforward algebra, the discretized optimization problem can be cast in the standard form of linear programming problems:
	\begin{align}
	\text{optimize}&~~F(c_{\vec v_1}..., c_{\vec v_n})=\sum_{i=1}^n c_{\vec v_i} N^{(A)}_i\;,\\
	\text{subject to}&~~\sum_{j=1}^{p+q+n}{\cal M}_{\alpha j} d_j  = b_\alpha, ~~ \alpha=1, ..., p+q+1\;, \label{eq:standard_form_1}\\
	\text{and}&~~ d_i \geq 0, ~~ i=1,..., n+p+q\;. \label{eq:standard_form_2}
	\end{align}
Here $F(c_{\vec v_1}..., c_{\vec v_n})$ is identified with the objective function, which depends on the ``decision variables'', $c_{\vec{v}_i}$, $i=1, ..., n$, which correspond in this case to the weights of the streams. Besides, $d_i$ are the components of the $(p+q+n)$-dimensional vector $(c_{\vec{v}_1},... c_{\vec{v}_n}, s_1, ... s_p, s'_1, ... s'_q)$, which contains, in addition of the ``decision variables'', a set of $p$ ``slack variables'', $s_\alpha$, and $q$ ``surplus variables'', $s'_\alpha$, which are introduced to cast the inequality constraints Eqs.~(\ref{eq:max_dis2},\ref{eq:min_dis2}) in the form of an equality constraint, as in Eq.~(\ref{eq:standard_form_1}). In the latter equation, ${\cal M}$ is a $(p+q+n)\times (p+q+1)$ matrix which explicitly reads
\begin{align}
{\cal M}=
\left(
\begin{array}{c|@{}c|@{}c}
  \begin{array}{ccc}
    1 & \cdots  &1  \\
  \end{array}
& 
\mathbf{0} 
& 
\mathbf{0} \\
\hline
\begin{array}{ccc}
  N_1^{(1)} &\cdots & N_n^{(1)} \\ 
  \vdots & \vdots & \vdots\\
  N_1^{(p)} &\cdots & N_n^{(p)} \\ 
\end{array} 
& 
\begin{array}{ccc}
    1  & \cdots & 0 \\ 
    \vdots & \ddots &  \vdots \\
    0 &  \cdots  & 1\\
  \end{array} 
& 
\mathbf{0}\\
\hline
\begin{array}{ccc}
  N_1^{(p+1)} &\cdots & N_n^{(p+1)} \\ 
  \vdots & \vdots & \vdots\\
  N_1^{(p+q)} &\cdots & N_n^{(p+q)} \\ 
\end{array}
  & 
  \mathbf{0}
  & 
\begin{array}{ccc}
  -1  & \cdots & 0 \\ 
  \vdots & \ddots &  \vdots \\
  0 &  \cdots  & -1\\
\end{array} \\
\end{array}\right)\;,
\end{align}
and $b_i$ are the components of the $(p+q+1)$-dimensional vector $(1, N_{\rm max}^{(1)},..., N_{\rm max}^{(p)}, N_{\rm min}^{(p+1)},..., N_{\rm min}^{(p+q)})$.

As is well known, the values for $d_i$ that minimize the linear program lie at the vertices of the feasible region defined by Eqs.~(\ref{eq:standard_form_1},\ref{eq:standard_form_2}) (see {\it e.g.} \cite{Bertsimas:1997}). If the rows of the matrix ${\cal M}$ are linearly independent, then there are $p+q+1$ variables taking non-zero values (so-called ``basic variables''), while the remaining $n-1$ variables must vanish. One should note that slack and surplus variables can also be basic, therefore the solution does not necessarily contain $p+q+1$ non-vanishing decision variables. In fact, when $r$ of the constraints are not saturated (namely, when $r$ constraints are ``not active''), then there are $r$ corresponding slack or surplus variables which are basic, and hence only $p+q+1-r$ non-vanishing decision variables. We then conclude that the optimized velocity distribution consists on a superposition of $p+q+1-r$ streams, with $r$ the number of inequality constraints that are not saturated, and therefore contains a maximum of $p+q+1$ streams and a minimum of 1 stream.

This approach has a number of applications:
\begin{itemize}
\item[{\bf \sc i.}] {\bf Derive a halo-independent upper limit on the cross section from a set of null results.}

We consider the null search experiments $A$ and $B_{\alpha}$, $\alpha=1, ..., p$, with outcomes $N^{(A)}\leq N^{(A)}_{\rm max}$, $N^{(B_\alpha)}\leq N^{(B_\alpha)}_{\rm max}$. We calculate, for a fixed dark matter mass and interaction cross section, the minimal outcome at the experiment $A$ from varying the velocity distribution, ${\rm min}\{N^{(A)}\}(\sigma,\text{m}_{\text{DM}})$, subject to the constraints from the set of $p$ null results from the experiments $B_{\alpha}$. The set of parameters $m_{\rm DM}$ and $\sigma$ is ruled out by the combination of the  upper limits from the $p$ experiments $B_{\alpha}$ {\it and} the upper limit from the experiment $A$ if
\begin{equation}
{\rm min}\{N^{(A)}\}(\sigma,\text{m}_{\text{DM}})~\geq~N^{(A)}_{\rm max}\;,
\label{eq:CondDDNT}
\end{equation}
from where one obtains a halo-independent upper limit on the scattering cross section as a function of the dark matter mass from combining the outcome of $p+1$ null results. 

We have for this case $p$ upper limit constraints, plus the equality constraint from normalization, therefore the optimized velocity distribution will consist of a superposition of streams, with a number that varies between $p+1$ (when all upper limits are saturated) and 1 (when no upper limit is saturated).

\item[{\bf \sc ii.}] {\bf Confront a detection claim to a set of null results in a halo independent manner.}

We consider the experiment $A$, with outcome $N^{(A)}\leq N^{(A)}_{\rm max}$, and the experiments $B_{\alpha}$,  $\alpha=1,...,p$, with outcome 
$N^{(B_\alpha)}_{\rm min}\leq N^{(B_\alpha)}\leq N^{(B_\alpha)}_{\rm max}$ for $\alpha=1,...,q$, and $N^{(B_\alpha)}\leq N^{(B_\alpha)}_{\rm max}$ for  $\alpha=q+1,...,p$. Namely, the experiments $B_{\alpha}$, $\alpha=1,...,q$, report the detection of a signal, while the experiments $A$ and $B_{\alpha}$, $\alpha=q+1, ...,p$, report upper limits. We now minimize, with respect to the velocity distribution, the outcome of the null search experiment $A$ for a given value of the dark matter mass and cross section, ${\rm min}\{N^{(A)}\} (\sigma,\text{m}_{\text{DM}})$, subject to the constraints $N^{(B_\alpha)}\leq N^{(B_\alpha)}_{\rm max}$, $\alpha=1, ..., p$, and $N^{(B_\alpha)}\geq N^{(B_\alpha)}_{\rm min}$, $\alpha=1, ..., q$. The set of parameters $m_{\rm DM}$ and $\sigma$ is incompatible with the upper limits from the experiment $A$ and the $p$ experiments $B_{\alpha}$,  $\alpha=1,...,p$, {\it and} with the detection claim of the $q$ experiments $B_{\alpha}$, $\alpha=1,...,q$,  if
\begin{align}
{\rm min}\{N^{(A)}\}(\sigma,\text{m}_{\text{DM}})~\geq~N^{(A)}_{\rm max}\;.
\label{eq:Cond-confront}
\end{align}

We have for this case $p$ upper limit and $q$ lower limit constraints, plus the equality constraint from normalization. On general grounds, the optimized velocity distribution contains $p+q+1$ streams, however it is clear that $q$ of the weights must vanish, since the slack and the surplus variables corresponding to the upper and the lower bounds reported by {\it the same} experiment cannot be zero simultaneously (this would require to saturate at the same time the lower and the upper bound). Therefore, also for this case the velocity distribution must be a superposition of streams, with a number that varies between $p+1$ (when all upper limits are saturated) and 1 (when no upper limit is saturated).

\item[{\bf \sc iii.}] {\bf Assess, in a halo-independent manner, the prospects for detection in a projected experiment given a set of upper limits from current experiments.}

We consider the experiments $B_\alpha$, $\alpha=1, ..., p$, providing the null results  $N^{(B_\alpha)}\leq N^{(B_\alpha)}_{\rm max}$ and the projected experiment $A$, which can claim detection if the outcome of the experiment is larger than $N_{\rm {det}}^{(A)}$. We maximize the outcome of the experiment $A$ for a given value of the dark matter mass and interaction cross section, ${\rm max}\{N^{(A)}\}(\sigma,\text{m}_{\text{DM}})$, subject to the constraints $N^{(B_\alpha)}\leq N^{(B_\alpha)}_{\rm max}$, $\alpha=1,...,p$. The regions of the parameter space where dark matter will escape detection at the experiment $A$ are defined by the condition:
\begin{align}
{\rm max}\{N^{(A)}\}(\sigma,\text{m}_{\text{DM}})~\leq N_{\rm det}^{(A)}\;.
\end{align}
As for the application  {\bf \sc i}, also in this case the optimized velocity distribution will consist of a superposition of streams, with a number that varies between $p+1$ and 1.

Furthermore, the regions of the parameter space that will be ruled out in a halo independent manner, in case of no detection, can be derived along the lines of our application  {\bf \sc i}:
\begin{align}
{\rm min}\{N^{(A)}\}(\sigma,\text{m}_{\text{DM}})~\geq N_{\rm det}^{(A)}\;.
\end{align}

\end{itemize}

In the next section we will illustrate these three applications with concrete examples. 

\section{Applications}
\label{sec:applications}

\subsection{Upper limit on the cross section from combining two null results}
\label{sec:two-null-results}

\begin{figure}[t!]
\begin{center}
\hspace{-0.75cm}
\includegraphics[width=0.49\textwidth]{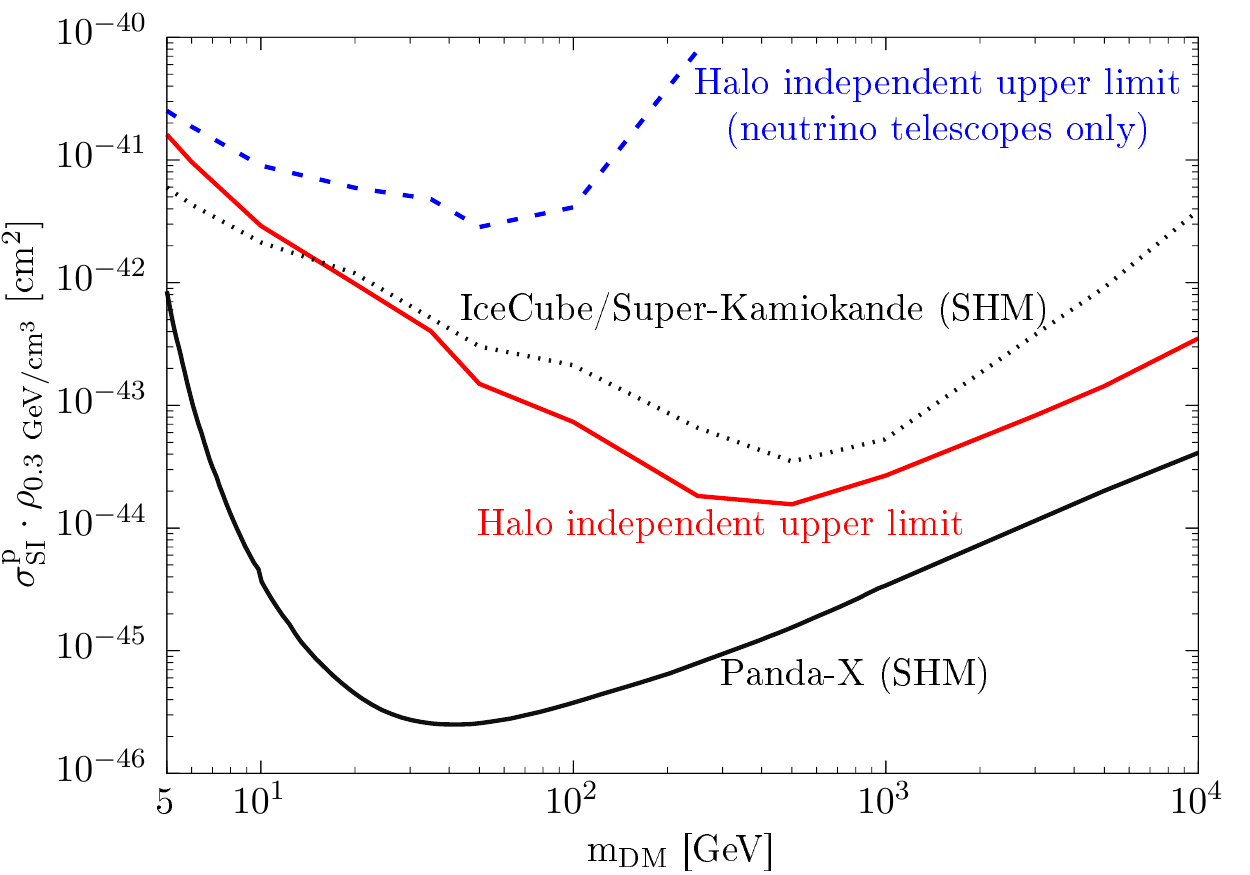}
\includegraphics[width=0.49\textwidth]{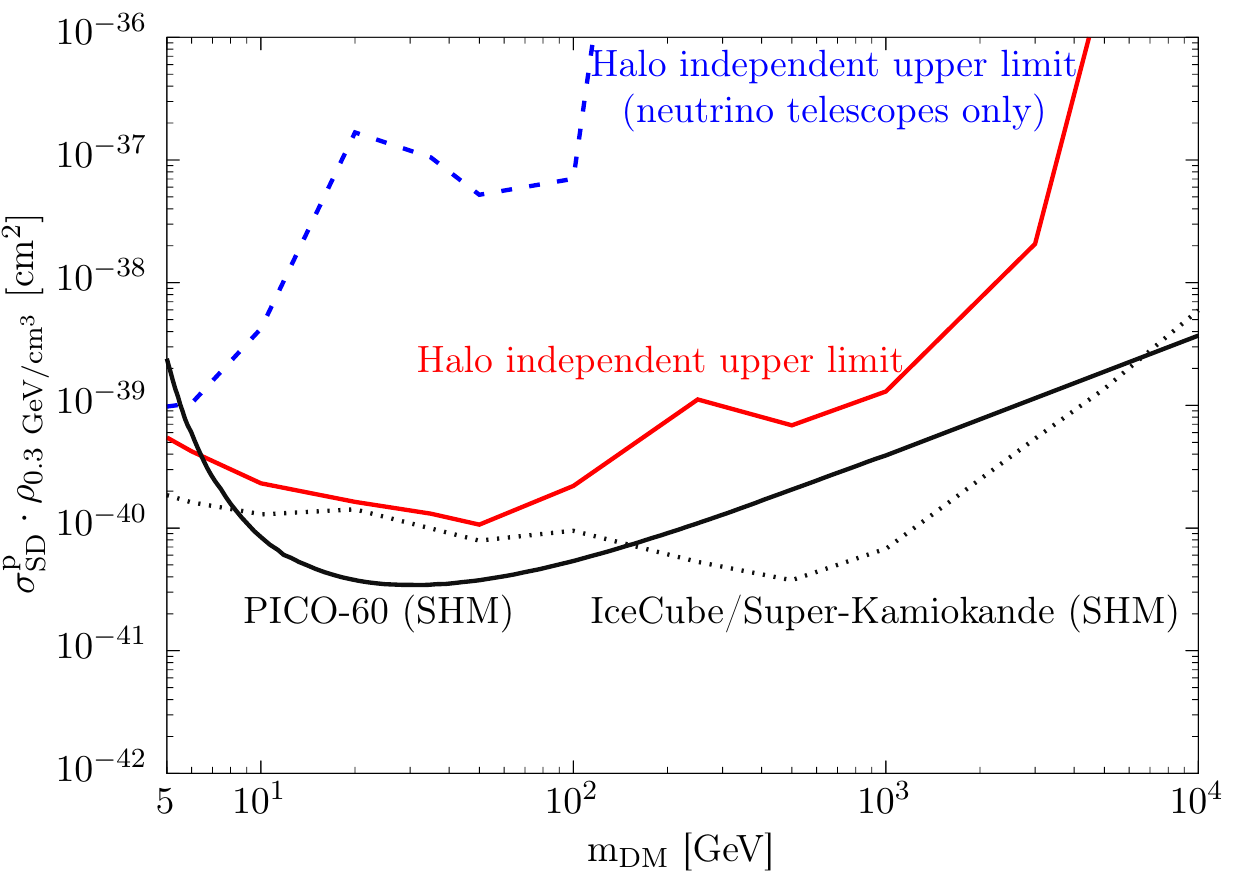}
\end{center}
\caption{Halo independent upper limits on the spin independent (left plot) and the spin dependent (right plot) scattering cross section as a function of the dark matter mass from combining the null results from PandaX (left plot) or PICO-60 (right plot) with those from the neutrino telescopes Super-Kamiokande and IceCube, as well as from neutrino telescopes only, assuming annihilation into $W^+W^-$ ($\tau^+\tau^-$ for $m_{\rm DM}<M_W$). We also show for comparison the upper limits reported by these experiments assuming the Standard Halo Model.}
\label{fig:ResultDDNT} 
\end{figure}

We illustrate this application of our method calculating a halo independent upper limit on the scattering cross section from combining the null results from a direct detection experiment and the null results from a neutrino telescope. Concretely, we will use the upper limit on the number of recoil events at PandaX Run 8~\cite{Tan:2016diz} and 9~\cite{Tan:2016zwf} for the SI interaction, or from PICO-60 \cite{Amole:2017dex} for the SD case, and the upper limit on the capture rate from IceCube, using 532 days of data,  \cite{Aartsen:2016zhm} and Super-Kamiokande, using 3903 days of data, \cite{Choi:2015ara}, assuming for concreteness annihilations into $W^+W^-$ ($\tau^+\tau^-$ for $m_{\rm DM}< M_W$). We note that the process of capture in the Sun does not depend on the characteristics of the neutrino telescope, therefore, for a given dark matter mass, it suffices to consider only the most stringent bound between the two. In our analysis, we will fix the time-dependent velocity of the Earth to the value giving the smallest recoil rate, such that our constraints can be regarded as conservative. Furthermore, with this prescription the initial problem of calculating the three-dimensional dark matter velocity distribution simplifies to a one-dimensional problem; for our implementation of the linear programming methods, we discretize the one-dimensional velocity space with 775 streams.\footnote{We have checked that increasing the resolution of the velocity space does not affect our conclusions.} We show the corresponding halo independent upper limits on the spin independent and spin dependent cross sections in Fig.~\ref{fig:ResultDDNT}.
 
We also show, as a dashed line, the halo independent upper limit on the scattering cross section from considering the null results from neutrino telescopes only, and which follows from the fact that neutrino telescopes probe the whole velocity space. More specifically, they correspond to the requirement $\min\{ C\}(\sigma,m_{\rm DM})\geq C_{\rm max}$. Details how to calculate $R_{\rm max}$ and $C_{\rm max}$ are provided in Appendix \ref{app:experiments}. The Figure also shows, for comparison, the upper limits published by the corresponding direct detection experiment or neutrino telescope, assuming the Standard Halo Model (SHM).  

The limits we obtain are remarkably strong and reach $\sigma_{\text{SI}}^{\text{p}}\lesssim 3\times 10^{-44}\,\text{cm}^{2}$ for the SI and $\sigma_{\text{SD}}^{\text{p}}\lesssim 10^{-39}\text{cm}^{2}$ for the SD scattering cross sections for $\text{m}_{\text{DM}}\sim 1~\text{TeV}$, assuming a local dark matter density $\rho_{\rm DM}=0.3\,{\rm GeV}/{\rm cm}^3$. We note that for very large dark matter masses it is not possible to derive a halo independent upper limit on the cross section. The reason is that, in this regime, capture in the Sun is possible only when the velocity in the Solar frame is  $v \leq \max_{r,i} \left\{v_{{\rm max},i}(r)\right\}\lesssim  {\rm max}_i\left\{2 v_{\rm esc}(0) \sqrt{m_{A_i}/m_{\DM}}\right\}$, where the maximum is taken over all possible distances to the center of the Sun and nuclei. On the other hand, nuclear recoils can be detected in a direct search experiment only when the velocity in the detector frame is $v^{\rm (D)}\gtrsim \min_{i} \left\{\sqrt{E_R/(2 m_{A_i})}\right\}$, which corresponds to a velocity in the Solar frame $v\gtrsim \min_{i} \left\{\left|v_{\oplus} -\sqrt{E_R/(2 m_{A_i})}\right|\right\}$; in this case the minimum must be taken over all nuclear species in the detector. Clearly, for sufficiently large dark matter masses, it is always possible to construct a velocity distribution consisting of streams with velocities which are too large to allow capture in the Sun and too small to produce a detectable nuclear recoil. These velocity distributions produce no signal in the direct detection experiment nor in the neutrino telescope and are therefore unconstrained. Concretely, for our concrete example the maximum mass that can be probed with our halo independent approach assuming scattering via the SI interaction only is $m_{\rm DM}\sim 165$ TeV  from combining PandaX and IceCube,~\footnote{Such large dark matter masses are, on the other hand, in tension with the unitarity limit for thermally produced dark matter~\cite{Griest:1989wd} and are contrived theoretically.} and assuming the SD interaction only,  $m_{\rm DM}\sim 4.5$ TeV from combining PICO-60 and IceCube.

For this specific application it is possible to determine analytically, for a given dark matter mass and cross section, the smallest possible scattering rate at a direct detection experiment compatible with the null results from a neutrino telescope, as well as the requirement that the velocity distribution is normalized to unity (for details, see Appendix \ref{app:analytical}). For a given dark matter mass, if the cross section is sufficiently small, the predicted capture rate for any stream configuration will be smaller than the upper limit from neutrino telescopes. Therefore, as the upper limit constraint cannot be saturated, there is only one non-vanishing decision variable, hence the optimized velocity distribution consists on just one stream. Clearly, the stream that produces the smallest recoil rate has zero velocity and accordingly the minimum recoil rate is zero. As the cross section increases, the upper limit constraint is saturated, the corresponding slack variable vanishes, and therefore there are two non-vanishing decision variables and the optimized velocity distribution consists of two streams.\footnote{We are grateful to Bradley Kavanagh for discussions about this point.}  In this part of the parameter space we find two possible solutions to the minimization problem:
\begin{align}
\boxed{\rm I}~~&f_{\rm I}=\frac{1}{2}\delta\Big(v-(\hat v-\epsilon)\Big)+\frac{1}{2}\delta \Big(v-(\hat v+\epsilon)\Big)\;, \nonumber\\
&{\rm with~}\epsilon\rightarrow 0{~\rm and~}  \hat v~ {\rm defined~by}\nonumber\\
&~~~~~C_{\hat v}=C_{\rm max}\;, \nonumber\\
& {\rm giving~ a ~rate}\nonumber\\
& ~~~~~R_{\rm I}=R_{\hat v}\;.\\
\boxed{\rm II}~~ & f_{\rm II}=\frac{C_{\rm max}}{C_{v_2}} \delta(v-v_1)+\left(1-\frac{C_{\rm max}}{C_{v_2}}\right)  \delta(v-v_{\rm max})\;,\nonumber\\
&{\rm with~} v_1 {\rm ~defined~by}\nonumber \\
&~~~~~\frac{d}{dv}\left[\frac{C_{\rm max}}{C_{v}} R_{v}+\left(1-\frac{C_{\rm max}}{C_{v}}\right) R_{v_{\rm max}}\right]\Big|_{v=v_1}=0\;, \nonumber\\
&{\rm giving~a~rate}\nonumber\\
&~~~~~R_{\rm II}=\frac{C_{\rm max}}{C_{v_1}} R_{v_1}+\left(1-\frac{C_{\rm max}}{C_{v_1}}\right) R_{v_{\rm max}}\;. 
\end{align}	
The optimized velocity distribution depends on the point in the parameter space and corresponds to the one giving the minimum between  $R_{\rm I}$ and $R_{\rm II}$. Finally, for a given dark matter mass, the  regions of the parameter space which are incompatible with the null search results from the given direct detection experiment and neutrino telescope are determined from
\begin{align}
\min\{R_{\rm I}, R_{\rm II}\}(\sigma,m_{\rm DM})\geq R_{\rm max}\;.
\end{align}

The region of the parameter space which is excluded using null results from neutrino telescopes only (bounded in Fig.~\ref{fig:ResultDDNT} with a dashed line), can also be determined analytically. The velocity distribution that minimizes the capture rate in the Sun, with the only constraint that the velocity distribution is normalized to unity, consists of a single stream with velocity $v_0$. This velocity can be calculated from $\partial C_v/\partial v\Big|_{v=v_0}=0$ which, due to the fact that the capture rate for streams decreases monotonically with the velocity, is achieved for $v_0=v_{\rm max}$. The excluded region is then defined by 
$\min \{C\}(\sigma,m_{\rm DM})= C_{v_{\rm max}}(\sigma,m_{\rm DM})\geq C_{\rm max}$. 

\subsection{DAMA confronted to null results}

The DAMA collaboration has reported a non-zero annual modulation of the scintillation light in sodium iodine detectors. The modulation has been consistently observed over 14 annual cycles, with a combined significance of 9.3$\sigma$~\cite{Bernabei:2013xsa}. Concretely, the modulation signal, as defined in Eq.~(\ref{eq:modulation-DAMA}), measured in the energy bins [2.0, 2.5], [2.5, 3.0] and [3.0, 3.5] keV are, respectively,  $(1.75\pm 0.37)\times 10^{-2}$, $(2.51 \pm 0.40)\times 10^{-2}$ and $(2.16 \pm 0.40)\times 10^{-2}~\text{day}^{-1}\,\text{kg}^{-1} \, \text{keV}^{-1}$. It has been claimed that the measured annual modulation could be explained from the time-dependent rate of scatterings of dark matter particles with the nuclei in the detector, which results from the changing alignment of the Earth's velocity with respect to the Sun's over the course of the year~\cite{Bernabei:2013xsa}. However, assuming the Standard Halo Model, the values of the dark matter parameters necessary to explain the measured modulation signal are in conflict with various null results, for instance, with the upper limits on the SI cross section from PandaX~\cite{Tan:2016diz}, IceCube~\cite{Aartsen:2016zhm} and Super-Kamiokande~\cite{Choi:2015ara}, and with the upper limits on the SD cross section from PICO-60~\cite{Amole:2017dex}, IceCube~\cite{Aartsen:2016zhm} and Super-Kamiokande~ \cite{Choi:2015ara}, among other experiments (a similar conclusion holds when considering the most general effective interaction compatible with the Galilean symmetry~\cite{Catena:2016hoj}).

Here, we apply our method to investigate, in a halo independent manner, the compatibility of the modulation signal reported by DAMA with the null results from direct detection experiments and/or from neutrino telescopes; in our analysis we will adopt as quenching factors $\text{Q}_{\text{Na}}=0.30$ and $\text{Q}_{\text{I}}=0.09$, as used by the DAMA collaboration in \cite{Bernabei:1996vj}, and we include the effects of channeling (and dechanneling) as described in \cite{Bozorgnia:2010xy}, adopting the largest values for the channeling fractions. For a fixed dark matter mass and cross section, we discretize the three-dimensional velocity space, expressed in galactic coordinates, in 100000 streams, and we calculate for each of the streams the modulation signal at DAMA, as defined in Eq.~(\ref{eq:modulation-DAMA}), in the  energy bins [2.0, 2.5], [2.5, 3.0] and [3.0, 3.5] keV, the scattering rate at PandaX (PICO-60) assuming the SI (SD) interaction only, and the capture rate inside the Sun. For the scattering rate at PandaX and PICO-60, we take properly into account the motion of the Earth around the Sun during the period of data taking, discretizing the orbit of the Earth in time intervals of 2 weeks. Finally, we implement the linear programming method, as described in Section~\ref{sec:optimization}

\begin{figure}[t!]
	\begin{center}
		\hspace{-0.75cm}
		\includegraphics[width=0.49\textwidth]{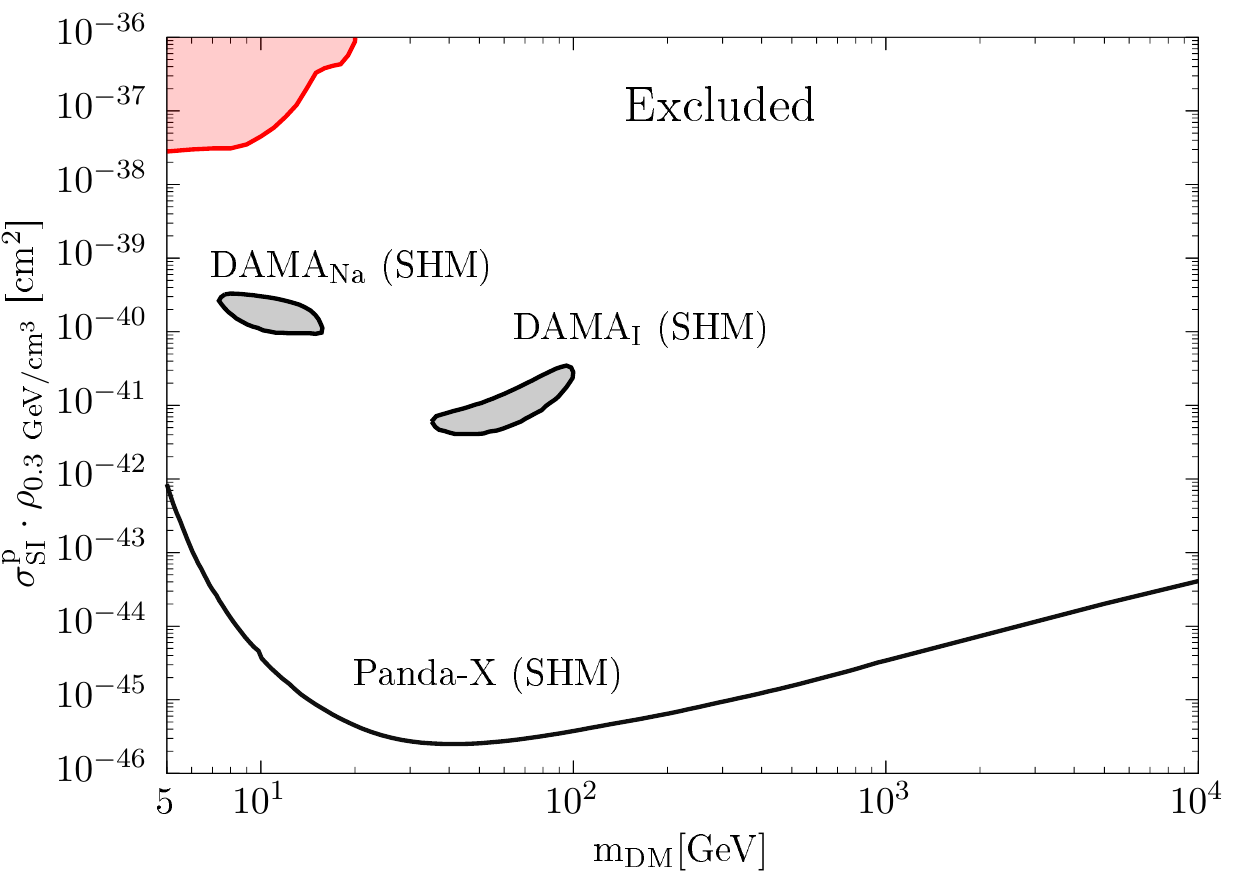}
		\includegraphics[width=0.49\textwidth]{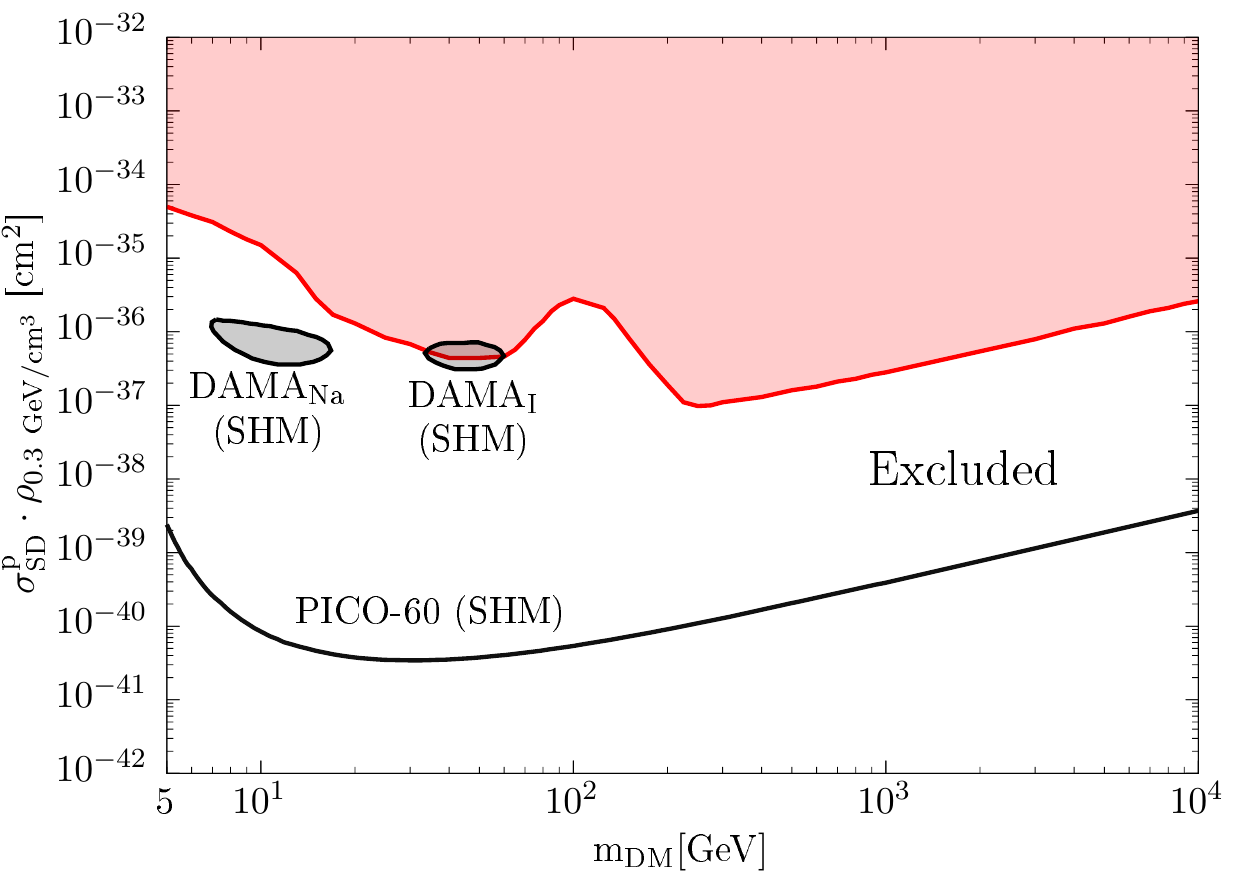} \\
		\hspace{-0.75cm}
		\includegraphics[width=0.49\textwidth]{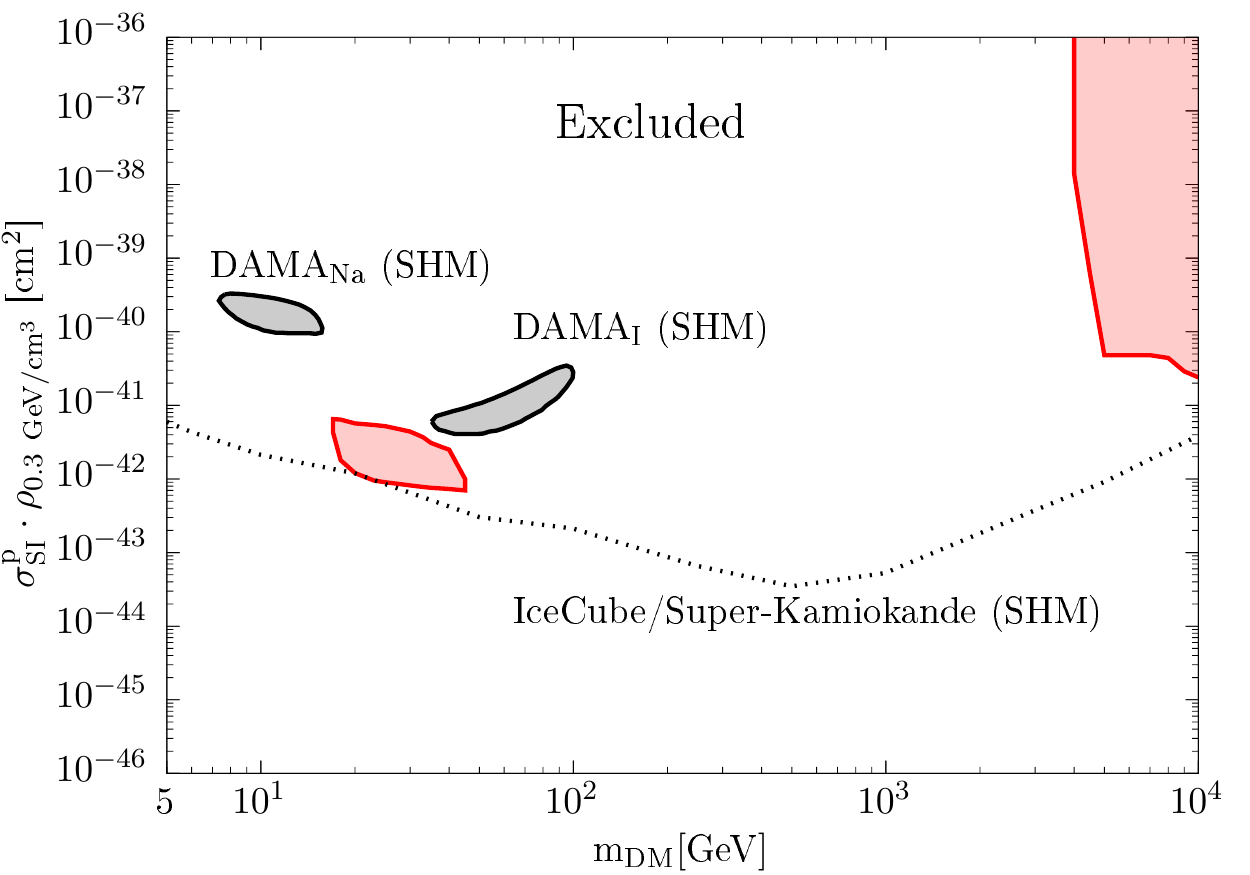}
		\includegraphics[width=0.49\textwidth]{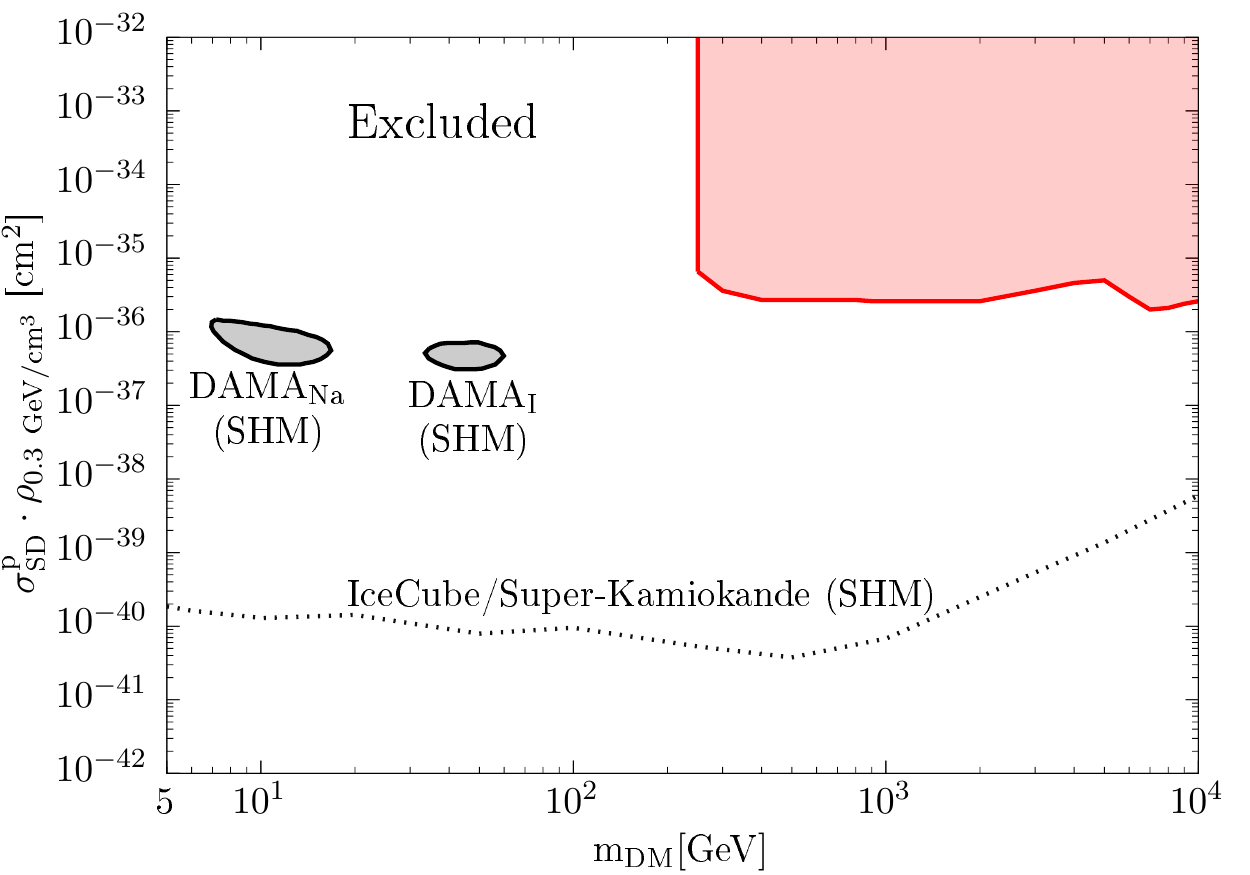}
	\end{center}
	\caption{Values of the spin-independent (left panels) or the spin-dependent (right panels) dark matter interaction cross section as a function of the mass leading to the modulation signal reported by DAMA, and which are compatible with the null searches from PandaX (upper left panel), PICO-60 (upper right panel) and the neutrino telescopes Super-Kamiokande and IceCube (lower panels). The gray region shows the parameters leading to the DAMA modulation signal, and the black line shows the upper limit on the cross section from the corresponding experiment, assuming the Standard Halo Model. }
	\label{fig:DAMA-x-section}
\end{figure}

In Fig.~\ref{fig:DAMA-x-section}, upper panels, we show as a white region the values for the SI (left plot) and SD (right plot) cross sections for which the measured values of $S^{\rm (DAMA)}_{[2.0,2.5]}$,  $S^{\rm (DAMA)}_{[2.5,3.0]}$ and  $S^{\rm (DAMA)}_{[3.0,3.5]}$ are incompatible with  the null results from PandaX or from PICO-60, respectively. The regions shown in pink, on the other hand, are compatible with the direct detection experiments, given our set-up. Some parts of this region, on the other hand, may be also excluded when taking into account in the analysis other constraints. For instance, the DAMA experiment has observed an annual modulation following approximately a cosine function, with a maximum in late May and a minimum in late November. However, our maximized solutions consist of a small number of streams, hence the time dependence of the recoil rate is expected to be more complex than a cosine function (see {\it e.g.} \cite{Freese:2012xd}), with a maximal rate not necessarily expected at June 1st, nor a minimal rate at December 1st, and with an amplitude different to the one given by the definition of the modulaton signal Eq.~(\ref{eq:modulation-DAMA}). Therefore, we expect that, if one adds to the set of constraints the requirement of reproducing the measured modulation rate in different time bins (or, alternatively, the requirement of reproducing the different coefficients in the Fourier series), then the excluded region will become larger. In this paper, and in order to illustrate our method, we just apply the simple test of reproducing the modulation signal, as defined in Eq.~(\ref{eq:modulation-DAMA}), in the three energy bins under consideration, deferring more detailed analyses for future work. Finally, we also show in the plots for comparison the upper bounds on the cross section reported by the corresponding direct detection experiments, as well as the regions favored by DAMA for the SI and SD interaction, taken respectively from \cite{Xu:2015wha} and from \cite{DelNobile:2015lxa}, all assuming the Standard Halo Model.

It is notable that there exist velocity distributions for which the DAMA signal is compatible with the null results from PandaX and PICO-60.\footnote{This conclusion, on the other hand, heavily relies on the inclusion of channeling effects in the calculation of the event rate, as the streams could be aligned with the axes of the DAMA crystals, hence enhancing the signal rate at this experiment.	When these effects are neglected, we found no region in the parameter space compatible with the PandaX and PICO-60 constraints.} Following the general discussion in Section \ref{sec:optimization}, the velocity distribution that minimizes the rate at a direct detection experiment subject to the constraint of reproducing the modulation signal in three energy bins as reported by DAMA, and subject to the normalization constraint, consists of the superposition of a maximum of four streams. In Table \ref{tab:distributions-DAMA-DD} we show the characteristics of the streams conforming the velocity distribution that minimizes the rates at PandaX and PICO-60, while reproducing the DAMA modulation signal, taking as benchmark values for the dark matter parameters $m_{\rm DM}=10\,{\rm GeV}$ and $\sigma_{\rm SI}=10^{-37}\,{\rm cm}^2$ for the former, and $m_{\rm DM}=3000\,{\rm GeV}$ and $\sigma_{\rm SD}=10^{-34}\,{\rm cm}^2$  for the latter.

The velocity distribution that minimizes the rate at PandaX consists of three streams with maximal velocities during the period of data taking, relative to the detector frame, equal to 257.1, 264.3 and 255.1 ${\rm km}/{\rm s}$. The streams $\#$1 and $\#$3 have maximal velocities smaller than the minimum velocity required to induce observable nuclear recoils for this DM mass, $v^{\,\rm (PandaX)}_{\rm min}=  259.6  \,{\rm km}/{\rm s}$, therefore, these streams produce no signal at PandaX. However, stream $\#$2 is above the threshold and can produce observable recoils. In this case, we obtain 0.0036, which is below the upper limit reported by PandaX. On the other hand, the three streams have velocities, both in June 1st and in December 1st, which are large enough to produce recoils in the DAMA experiment with energy between 3.0 and 3.5 keV, as the velocity threshold in this bin is $v^{\,\rm (DAMA[3.0,3.5])}_{\rm min}= 143.9 \,{\rm km}/{\rm s}$. Therefore, these streams can produce a modulation signal, as defined in Eq.~(\ref{eq:modulation-DAMA}), in the three energy bins, [2.0, 2.5], [2.5, 3.0] and [3.0, 3.5] keV; the weights and directions of the streams are such that the modulation signal in each of these three bins is reproduced. Similarly, the velocity distribution that minimizes the rate at PICO-60 in our benchmark point consists of four streams with maximal velocities in the detector frame equal to 112.4, 117.5, 107.3 and 94.1 km/s, which lie below the velocity threshold of PICO-60, $v^{\,\rm (PICO-60)}_{\rm min}=  120.0  \,{\rm km}/{\rm s}$, and hence do not produce observable recoils in this experiment. However, these streams have velocities in the DAMA detector frame above its threshold in the highest energetic bin, $v^{\,\rm (DAMA[3.0,3.5])}_{\rm min}= 20.3 \,{\rm km}/{\rm s}$, both in June and in December, and induce an annual modulation signal as observed by the experiment. 

We note that in both cases the optimized solutions contain streams with velocities very close to the threshold of the experiment. Namely, for the SI interaction, stream $\#2$ has a maximum velocity during the period of data taking slightly above the threshold and produces a non-zero number of events at PandaX, although below the measured upper limit, while for the SD interaction, stream $\#2$ has a maximum velocity slightly below the threshold and produces no event. Therefore, even small perturbations of these configurations may induce a number of recoil events in excess of the upper limit reported by the experiments. More concretely if these two streams are smeared by a Gaussian distribution with width $\Delta v=1.7 \,{\rm km}/{\rm s}$ and $\Delta v=1.6 \,{\rm km}/{\rm s}$, respectively, these velocity distributions would be excluded.

As apparent from the figure, there exists a minimum value of the cross-section necessary to produce the DAMA modulation signal while circumventing the null results from PandaX and PICO-60. Our analysis also shows that there exist solutions for arbitrarily large cross sections. To elucidate this point, let us consider the optimized velocity distribution for our benchmark dark matter parameters $m_{\rm DM}=10\,{\rm GeV}$ and $\sigma_{\rm SI}=10^{-37}\,{\rm cm}^2$, assuming the SI interaction only, and which consists of three streams. Clearly, an increase in the cross section can be compensated with a decrease of the weights of the three streams, such that the rates at PandaX and DAMA remain the same. In doing this, however, the velocity distribution is no longer normalized to unity. On the other hand, this can be amended by introducing a fourth stream, with velocity below the thresholds of both experiments, and with a weight chosen to fulfill the normalization condition. Therefore, it is always possible to construct a velocity distribution that compensates the increase in the cross section and which produces the same number of recoil events. One should also note that as the cross section increases, a smaller and smaller spread in the streams is required in order not to produce a number of recoil events in PandaX in excess with observations, and the solution becomes accordingly more and more tuned.

\begin{table}[t!]
	\begin{center}
		\begin{tabular}{c|cccc}
			& stream \#1 & stream \#2 & stream \#3 & stream \#4 \\
			\hline 
			$c_{\vec v_i}$ &  0.54 &  0.28 & 0.18  & $-$ \\
			$\vec v_i~ [{\rm km}/{\rm s}]$ &$(-10,-123,191)$  & $(100,-167,-161)$ &  $(56,119,-183)$  & $-$ \\
			$|\vec{v}_{i\,,{\rm max}}^\text{~(PandaX)}|~ [{\rm km}/{\rm s}]$ & 257.1  & 264.3 & 255.1 & $-$\\
			$|\vec{v}_{i,{\rm June}}^{~\rm (DAMA)}|~ [{\rm km}/{\rm s}]$ & 256.1 & 245.0 & 195.7  & $-$\\
			$|\vec{v}_{i,{\rm Dec}}^{~\rm (DAMA)}|~ [{\rm km}/{\rm s}]$ & 198.8  & 263.2 & 255.1 & $-$\\
			\hline
			$c_{\vec v_i}$ & 0.35 & 0.34 & 0.17 & 0.14  \\
			$\vec v_i~ [{\rm km}/{\rm s}]$ & $(-91,10,20)$ &$(-1,100,2)$ & $(35,46,-52)$ & $(-37,-62,74)$ \\
			$|\vec{v}_{i\,,{\rm max}}^{~(\text{PICO})}|~ [{\rm km}/{\rm s}]$ & 112.4  & 117.5 & 107.3 & 94.1 \\
			$|\vec{v}_{i,{\rm June}}^{~\rm (DAMA)}|~ [{\rm km}/{\rm s}]$ & 109.9 & 89.6 & 49.0 & 133.0  \\
			$|\vec{v}_{i,{\rm Dec}}^{~\rm (DAMA)}|~ [{\rm km}/{\rm s}]$ & 85.9 & 117.3 & 107.0 & 74.1  \\
		\end{tabular}
	\end{center}
\caption{Velocities in galactic coordinates ($\vec v_i$) and weights ($c_{\vec v_i}$) of the streams of dark matter particles that conform the velocity distributions which reproduce the DAMA modulation signal in the energy bins [2.0, 2.5], [2.5, 3.0] and [3.0, 3.5] keV and are compatible with the null searches from PandaX assuming scatterings via the SI interaction only with $m_{\rm DM}=10\,{\rm GeV}$ and $\sigma_{\rm SI}=10^{-37}\,{\rm cm}^2$ (upper panel), or from PICO-60 assuming scattering via the SD interaction only with $m_{\rm DM}=3000\,{\rm GeV}$  and $\sigma_{\rm SD}=10^{-34}\,{\rm cm}^2$ (lower panel). $|\vec{v}_{i,{\rm max}}^{~\rm (EXP)}|$ denotes the maximal velocity of the corresponding stream in the frame of the experiment EXP over the period of data taking, and $|\vec{v}_{i,{\rm June}}^{~\rm (DAMA)}|$ and $|\vec{v}_{i,{\rm Dec}}^{~\rm (DAMA)}|$ are the velocities in the DAMA detector frame in June and December, respectively.
}
\label{tab:distributions-DAMA-DD}
\end{table}

 The bottom panels of Fig. \ref{fig:DAMA-x-section} show, on the other hand, the values of the SI and SD cross-sections where the DAMA claim is compatible with the null results from IceCube and Super-Kamiokande, assuming annihilations into $W^+W^-$. As before, we find velocity distributions for which the modulation signal reported by DAMA is compatible with the null results from the neutrino telescopes Super-Kamiokande and IceCube. The regions of the parameter space where such velocity distributions exist are shown in pink. Both for the SI and SD interactions, there exists an allowed region at high dark matter masses, concretely $m_{\rm DM}\gtrsim 4\,{\rm TeV}$ for the SI interaction and $m_{\rm DM}\gtrsim 250\,{\rm GeV}$ for the SD interaction. Moreover, assuming only the SI interaction, we find an allowed region for masses  $\sim 20-40$ GeV and cross sections $~10^{-42}-10^{-41}\,{\rm cm}^2$.

In this case the optimized velocity distribution also consists of a maximum of four streams; the characteristics of these streams are shown in table \ref{tab:distributions-DAMA-NT}, taking the same benchmark dark matter parameters as in table \ref{tab:distributions-DAMA-DD}: $m_{\rm DM}=10\,{\rm GeV}$ and $\sigma_{\rm SI}=10^{-37}\,{\rm cm}^2$ for the SI interaction, $m_{\rm DM}=3000\,{\rm GeV}$ and $\sigma_{\rm SD}=10^{-34}\,{\rm cm}^2$  for the SD interaction. The optimized velocity distribution, assuming only the SI interaction, consists of two streams with fairly high velocities, 728.0 and 734.2 km/s in the rest frame of the Sun, and one stream with lower velocity, 362.8 km/s, but with a very small weight. Dark matter particles in these three streams can be captured inside the Sun at a sufficiently large rate to achieve equilibration, and generate in annihilations a neutrino flux in excess with observations. For the benchmark point for the SD interaction, on the other hand, the dark matter particles in the three streams have a large kinetic energy, so that after scattering with one nucleus in the solar interior and transferring part of its energy, the DM particle still has a velocity which is larger than the escape velocity. Capture in this case is very inefficient and as a result these streams are untestable using neutrino telescopes.

For very large values of the cross section we find no allowed solutions, contrary to the behavior found when comparing DAMA with the null results from direct detection experiments (the excluded regions in the bottom panels of Fig. \ref{fig:DAMA-x-section} lie, however, outside of the Figure). Following a similar rationale as for the combination of DAMA with other direct detection experiments, an increase in the cross section can be compensated with a decrease in the weights of the streams reproducing the DAMA signal, whereas the normalization condition can be fulfilled by postulating an additional stream with velocity below the threshold of the experiment. Now, the dark matter particles in this low-velocity stream can be efficiently captured inside the Sun, due to the large cross section, and therefore produce a neutrino flux. As a result, reproducing the DAMA modulation signal with the constraints on the capture rate from neutrino telescopes, translates into a lower and an upper bound on the scattering cross section; the former, from requiring a large enough modulation signal, and the latter, from requiring a small enough capture rate. 

\begin{table}[t!]
	\begin{center}
		\begin{tabular}{c|cccc}
			& stream \#1 & stream \#2 & stream \#3 & stream \#4 \\
			\hline 
			$c_{\vec v_i}$ &  0.56 & 0.44 & $6\times 10^{-5}$ & $-$ \\
			$\vec v_i~ [{\rm km}/{\rm s}]$  & $(120,-623,-357)$ & $(110,-643,-337)$&  $(110,177,-297)$ &$-$ \\
			$|\vec{v}_i^{~\text{(Sun)}}|~ [{\rm km}/{\rm s}]$  & 728.0 & 734.2 & 362.8 & $-$\\
			$|\vec{v}_{i,{\rm June}}^{~\rm (DAMA)}|~ [{\rm km}/{\rm s}]$ & 728.0  & 735.4 & 333.0 & $-$\\
			$|\vec{v}_{i,{\rm Dec}}^{~\rm (DAMA)}|~ [{\rm km}/{\rm s}]$& 734.2 & 729.1  & 392.6 & $-$\\
			\hline
			$c_{\vec v_i}$ & 0.51 &0.47 & 0.02 &  \\
			$\vec v_i~ [{\rm km}/{\rm s}]$ & $(70,107,-167)$ & $(-60,-113,173)$& $(-50,-103,143)$   & $-$ \\
			$|\vec{v}_i^{~\text{(Sun)}}|~ [{\rm km}/{\rm s}]$ & 210.3 & 215.2 & 183.2 & $-$\\
			$|\vec{v}_{i,{\rm June}}^{~\rm (DAMA)}|~ [{\rm km}/{\rm s}]$& 180.5  & 244.9 & 212.9 & $-$\\
			$|\vec{v}_{i,{\rm June}}^{~\rm (DAMA)}|~ [{\rm km}/{\rm s}]$ & 240.1 & 185.4 & 153.5  & $-$
		\end{tabular}
	\end{center}
	\label{tab:distributions-DAMA-NT}
	\caption{Same as Table \ref{tab:distributions-DAMA-DD}, but for the dark matter streams that conform the velocity distributions which reproduce the DAMA modulation signal and are compatible with the null searches from the neutrino telescopes IceCube and Super-Kamiokande, assuming annihilations into $W^+W^-$ ($\tau^+\tau^-$ for $m_{\rm DM}<M_W$). Here, $|\vec{v}_i^{~\text{(Sun)}}|$ denotes the stream velocity in the solar frame.}
\end{table}

One should note that the velocity distribution that minimizes the recoil rate at a direct detection experiment and the one that minimizes the capture rate at the Sun are qualitatively different: in the former case, the velocity distribution should not contain streams with very high velocities (or have a small weight), in order to prevent a too large scattering rate at PandaX or PICO-60, and in the latter, the velocity distribution should not contain streams with very low velocities (or have a small weight) in order to prevent efficient capture inside the Sun. As the requirements for the velocity distributions are opposite, it is interesting to investigate whether there exist velocity distributions for which the DAMA modulation signal can be made compatible with the null results from other direct detection experiments {\it and} from the null results from neutrino telescopes. 

\begin{figure}[t!]
	\begin{center}
		\hspace{-0.75cm}		\includegraphics[width=0.49\textwidth]{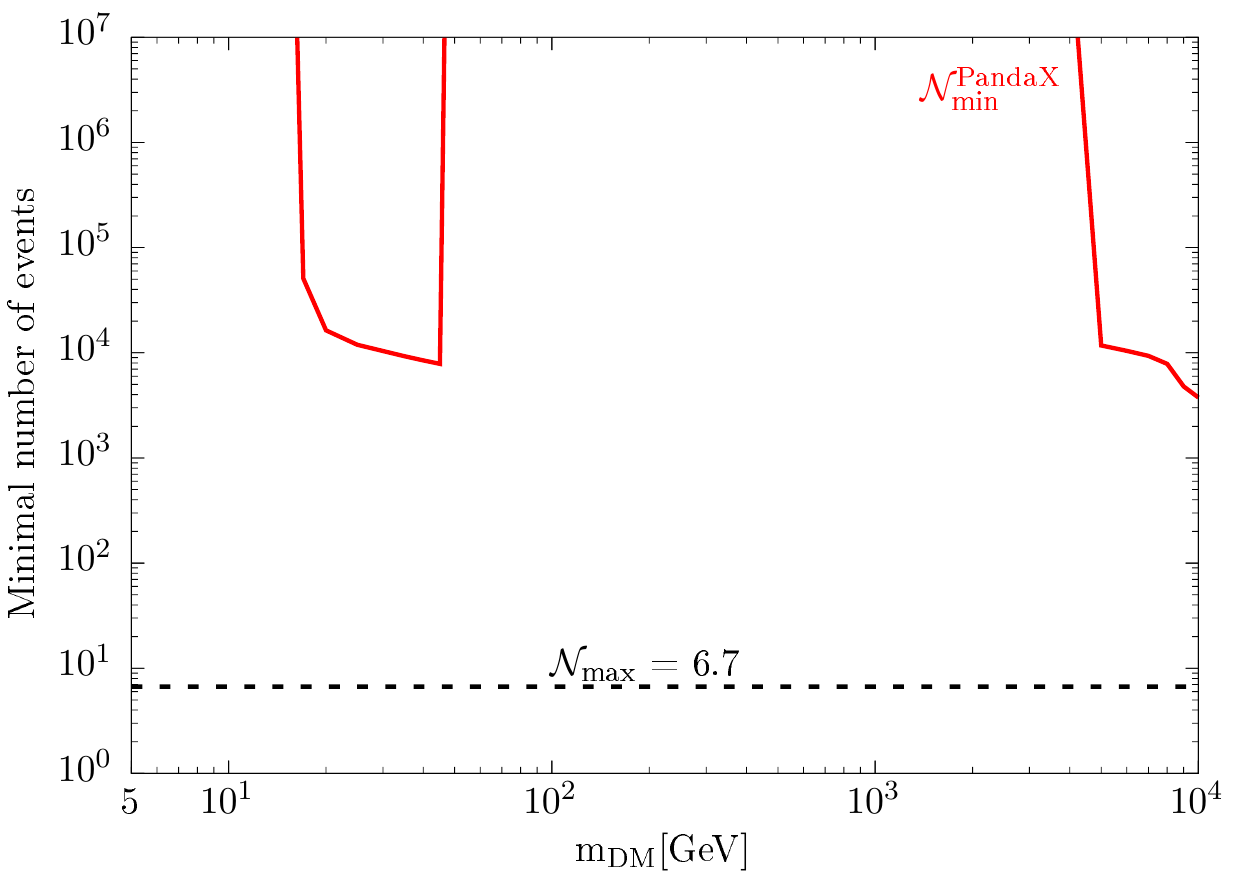}
		\includegraphics[width=0.49\textwidth]{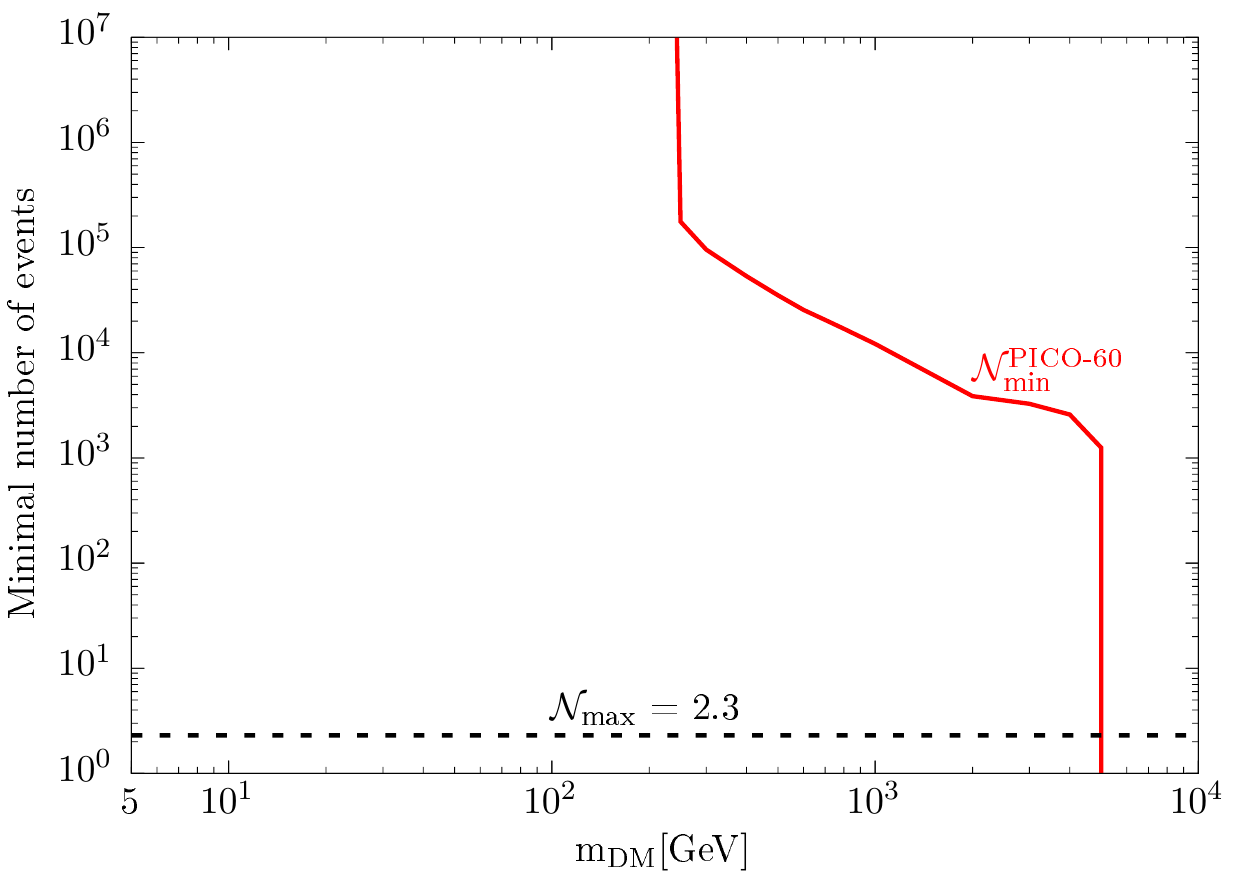}
	\end{center}
	\caption{Minimum value of the number of dark matter induced recoil events expected at PandaX (left panel) and PICO-60 (right panel), under the assumption that the modulation signal reported by DAMA is the result of dark matter scatterings induced by the SI interaction (left panel) or the SD interaction (right panel), and from requiring compatibility with the null search results from neutrino telescopes. Here, ${\cal N}_{\rm max}$ denotes the 90\% C.L.  upper limit on the number of recoil events from the corresponding experiment.}
	\label{fig:MaxSignal}
\end{figure}

We have applied our method to calculate the minimum number or recoil events at PandaX and PICO-60, with the constraints of correctly reproducing the DAMA modulation signal in the three relevant energy bins, the upper limit on the capture rate from neutrino telescopes, and the normalization condition (in this case, the optimized velocity distribution consists of a superposition of a maximum of five streams). The result is shown in Fig.\ref{fig:MaxSignal}. The plot shows that no dark matter particle interacting with nucleons via the SI interaction only can produce the observed signal at DAMA, while being compatible with the null searches from PandaX and from current neutrino telescopes, if the DM mass is in the range 5 GeV- 10 TeV, as the number of expected events at PandaX is in this mass range larger than $\sim 3000$. Only for very large dark matter masses, where capture in the Sun becomes very ineffective, it is possible to find velocity distributions where the DAMA signal is compatible with the null searches, as neutrino telescopes do not constrain this region of the parameter space. However, as discussed in Subsection \ref{sec:two-null-results}, these regions are in tension with the unitarity limit for thermally produced dark matter and are contrived theoretically. On the other hand, we find that if the DM scatters via the SD interaction only, there are velocity distributions where the experiments considered are compatible, if the DM mass is larger than $\sim 4.5$ TeV. However, and as discussed above, the velocity distributions for which DAMA is compatible with the null results from PandaX or PICO-60 require a very small velocity dispersion in the dark matter streams and can be regarded as fine tuned.

\subsection{Prospects for LZ from current null results}

Finally, we apply our halo independent method to assess the prospects to observe a dark matter signal in a future experiment, in view of null results from current experiments.  Concretely, we will identify the regions of the parameter space where a signal is expected at the projected LUX-ZEPLIN (LZ) experiment~\cite{Akerib:2015cja}, regardless of the velocity distribution, the regions which will remain untested, and the regions which may produce signals, depending on the velocity distribution. To identify all these regions we will use the null search results from  SuperCDMS\footnote{The halo independent comparison is only meaningful among experiments with different characteristics {\it e.g.}, different energy thresholds or different energy resolutions. Therefore,  to illustrate our method and assess the reach of LZ, which is based on a xenon target, we will employ the null results from SuperCDMS, which is based on a germanium target, and which has very different characteristics than LZ.}~\cite{Agnese:2014aze} and/or from the neutrino telescopes IceCube~\cite{Aartsen:2016zhm}  and Super-Kamiokande~\cite{Choi:2015ara}, when analyzing scattering induced by the SI interaction only, and from PICO-60 \cite{Amole:2017dex} and/or from IceCube and Super-Kamiokande, when considering the SD interaction only. As in the rest of this work, we will assume equilibration between dark matter capture and annihilation in the solar interior, and that dark matter annihilates only into $W^+W^-$ (or $\tau^+\tau^-$ for $m_{\rm DM}<M_W$).  Similarly to the numerical implementation of the method presented in Subsection \ref{sec:two-null-results}, we 
will fix the time-dependent velocity of the Earth to the value giving the smallest (largest) recoil rate over the year, in order to conservatively define the regions which will be tested (remain untestable) at LZ, and in order to reduce the three-dimensional problem of calculating the optimized velocity distribution into a simpler one-dimensional problem; in our calculation we discretized the one-dimensional velocity distribution with 775 streams.

\begin{figure}[t!]
\begin{center}
\hspace{-0.75cm}
\includegraphics[width=0.49\textwidth]{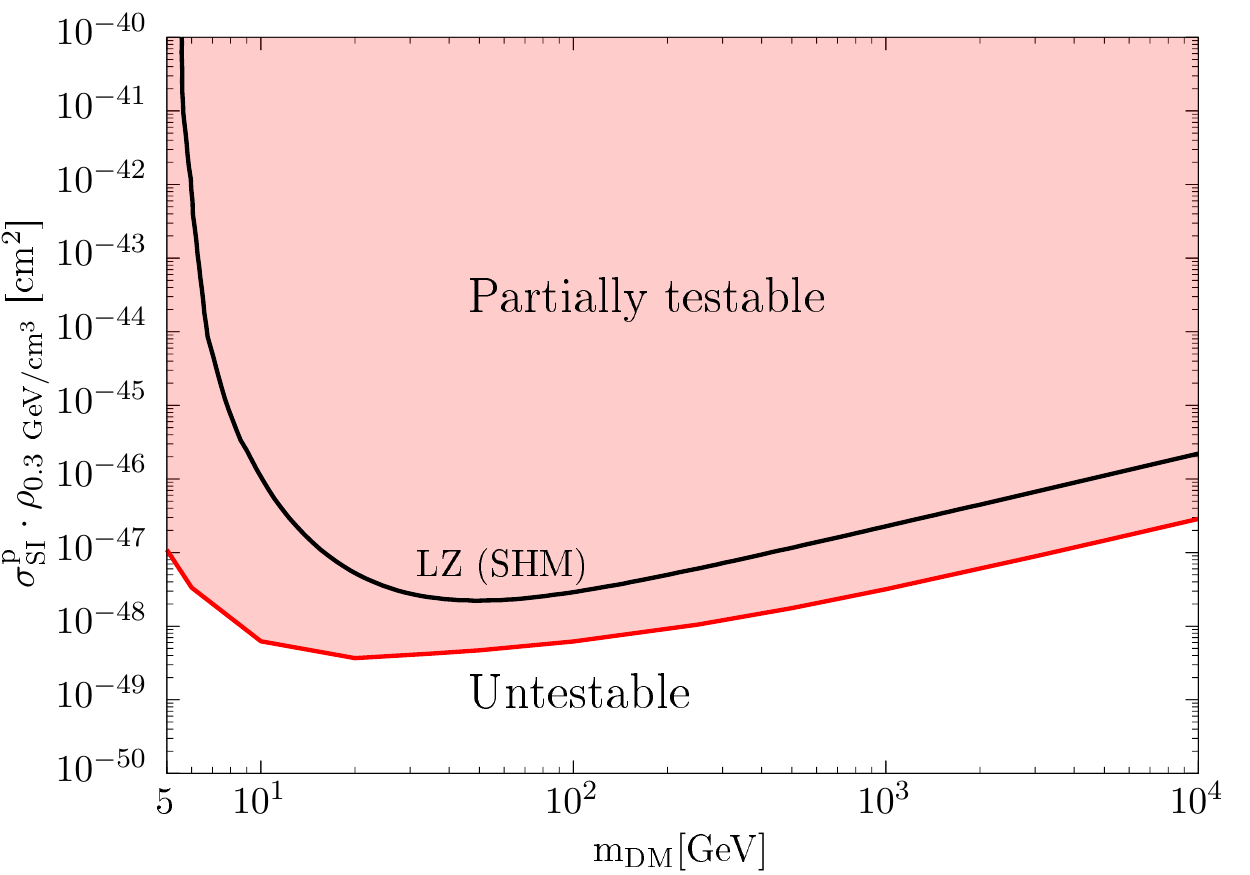}
\includegraphics[width=0.49\textwidth]{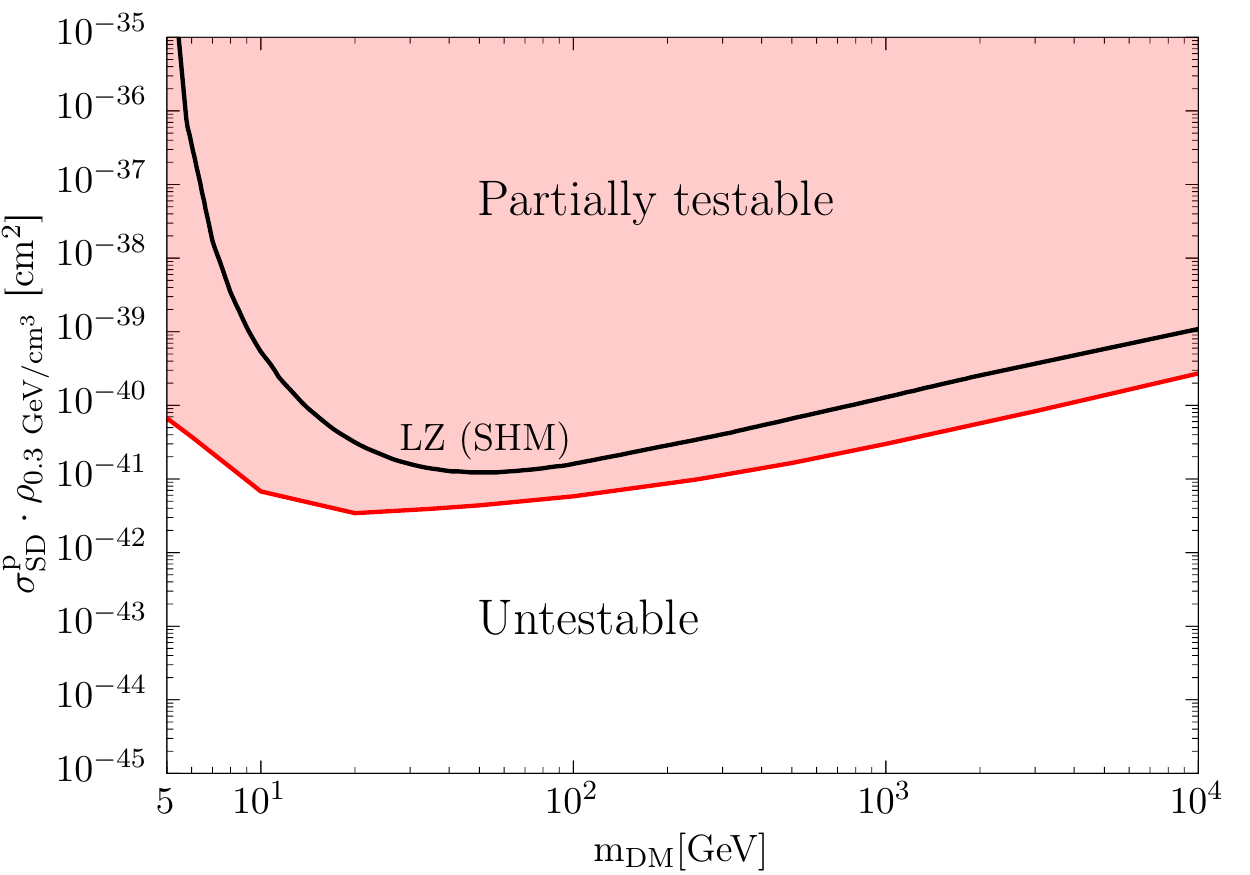}\\
\hspace{-0.75cm}
\includegraphics[width=0.49\textwidth]{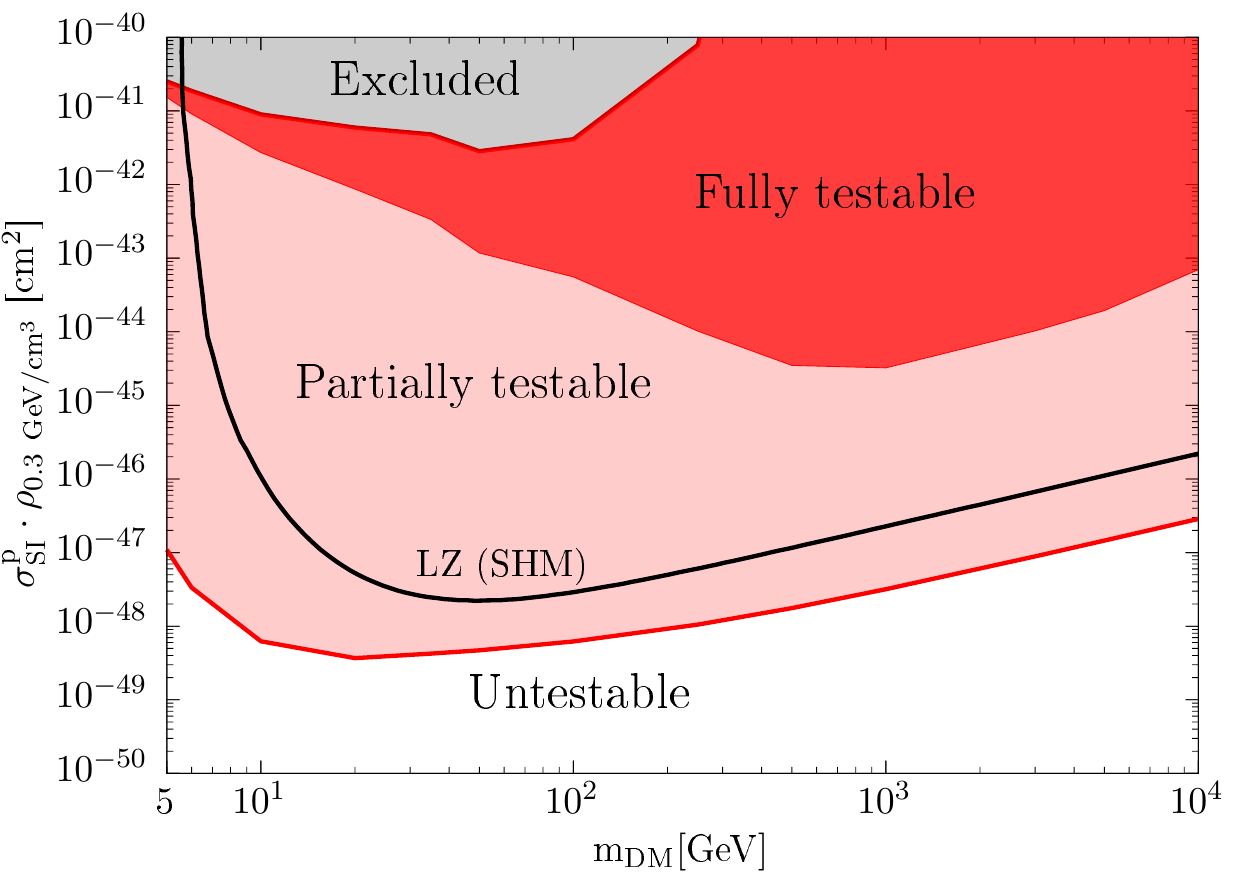}
\includegraphics[width=0.49\textwidth]{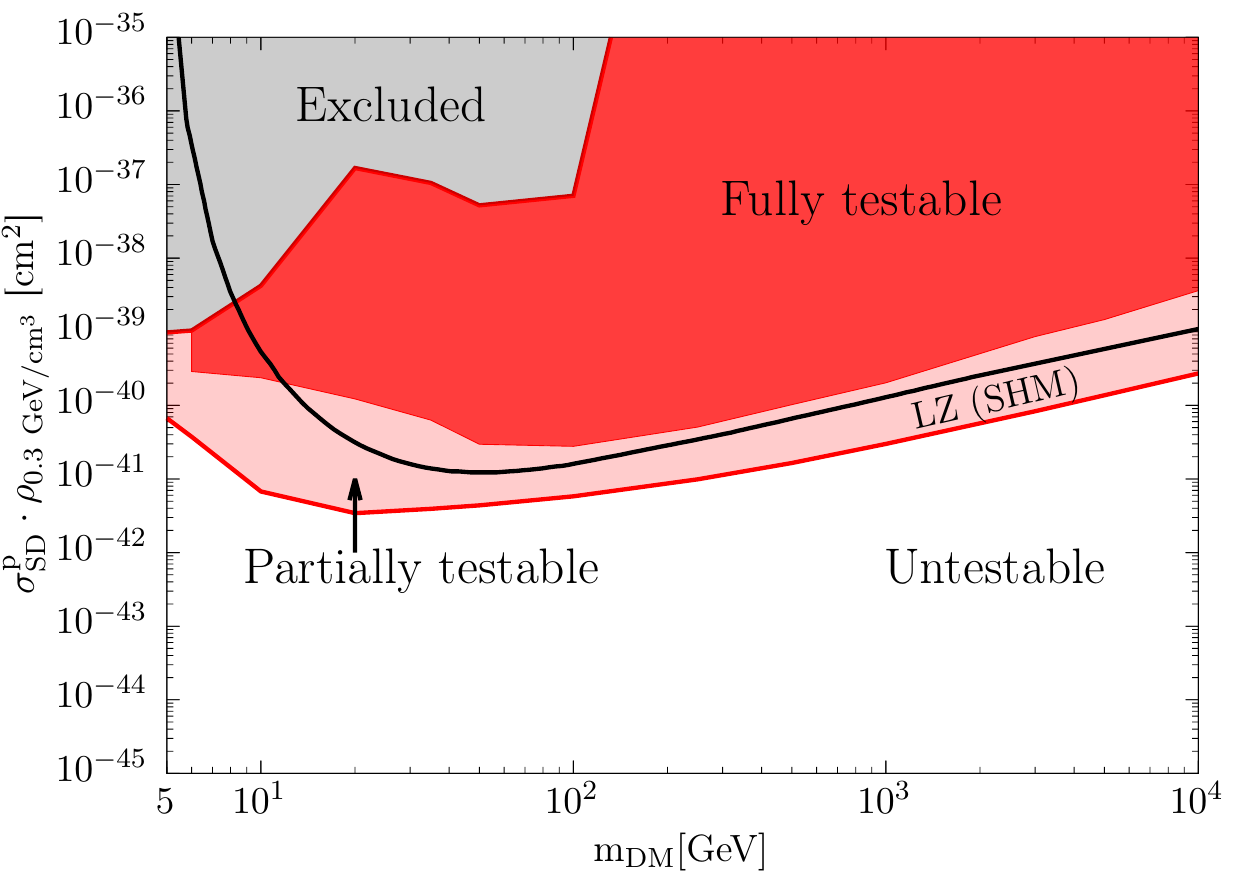}\\
\hspace{-0.75cm}
\includegraphics[width=0.49\textwidth]{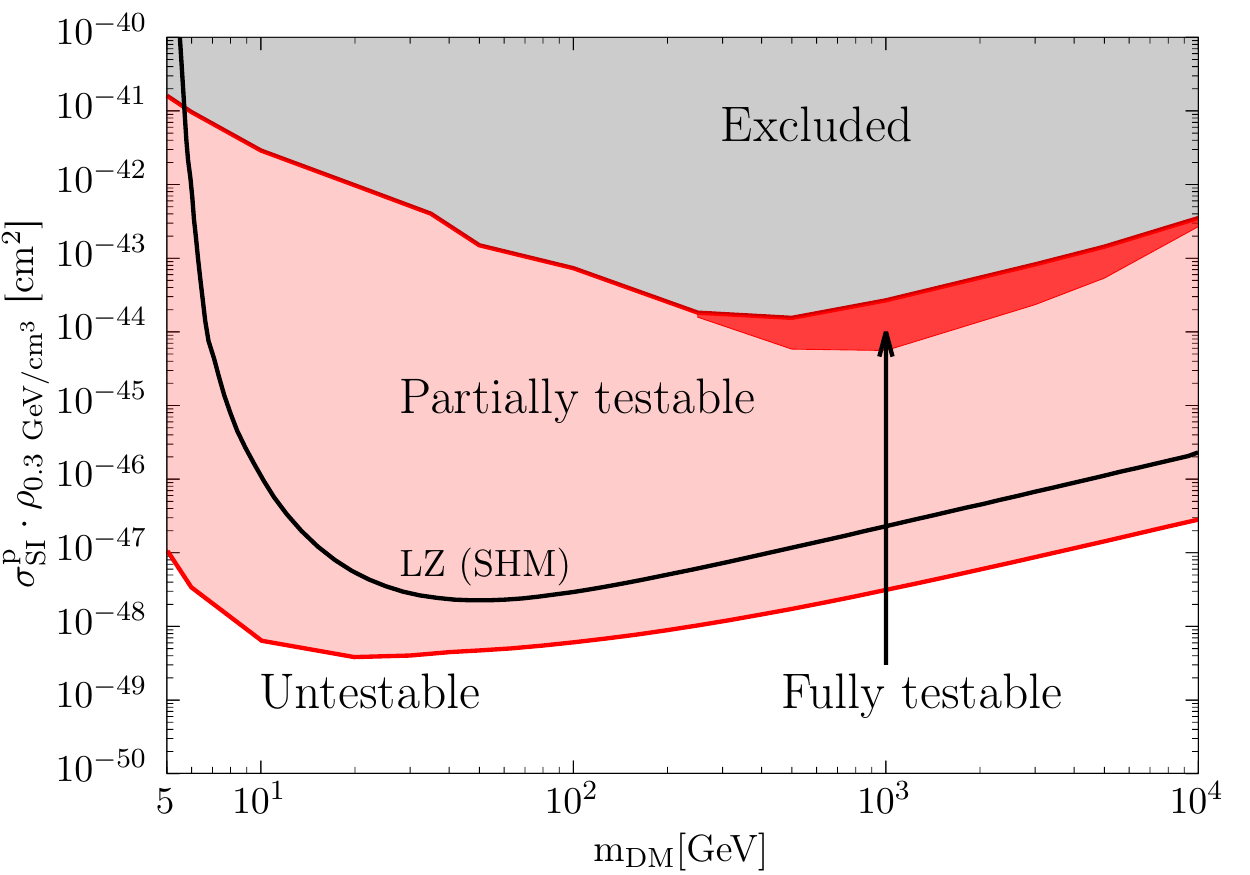}
\includegraphics[width=0.49\textwidth]{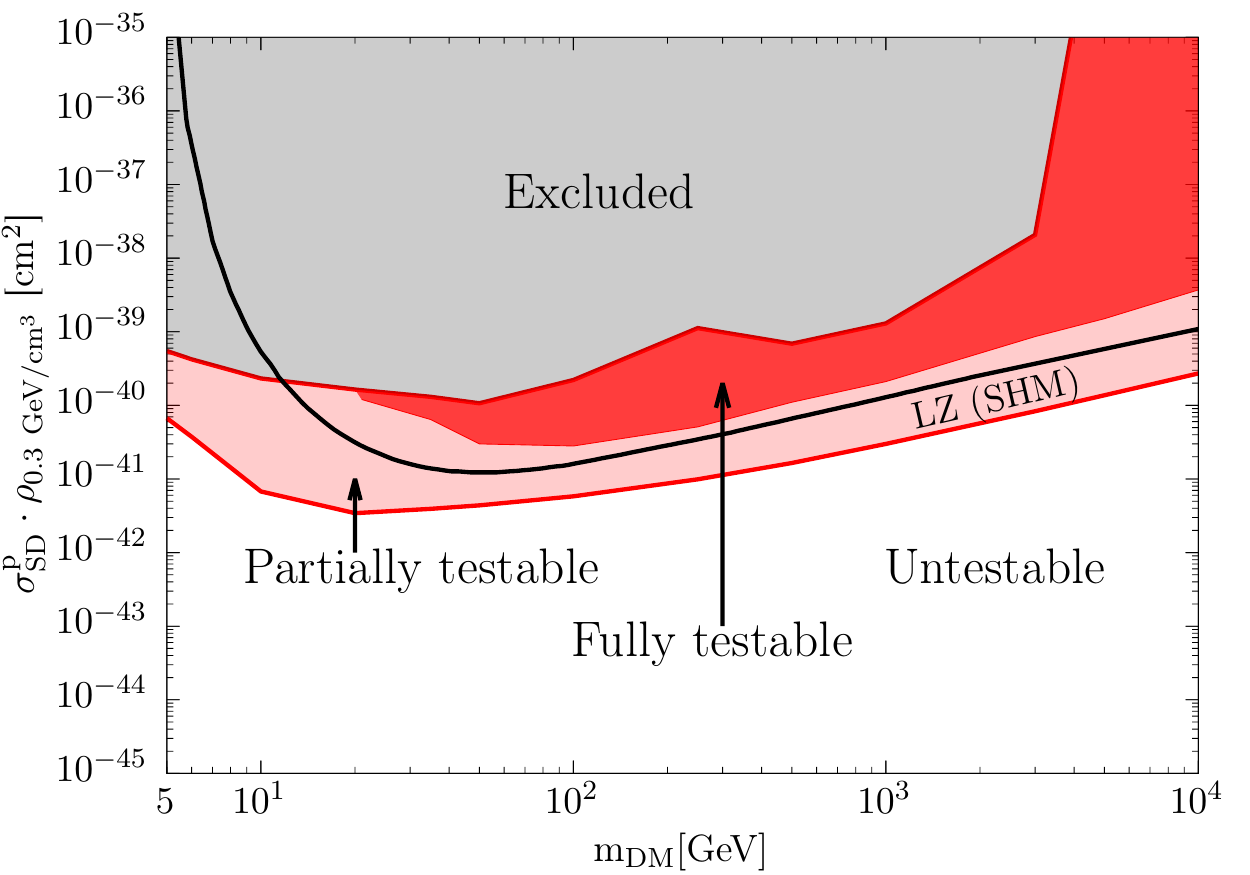}\\
\end{center}
\caption{Values of the SI cross section (left panels) or SD cross section (right panels) where the maximum number of events expected at LZ is smaller than 1 for all velocity distributions (white regions), is larger than 1 for at least one velocity distribution (pink regions) or is larger than 1 for all velocity distributions (red regions) in view of the current null search results from the direct detection experiments SuperCDMS or PICO-60 (upper panels), from neutrino telescopes (middle panels), or from considering both search strategies simultaneously (lower panels). The gray regions in the middle and bottom plots are ruled out, in a halo independent manner, by current null results from neutrino telescopes only, and from combining null results from neutrino telescopes and direct detection experiments, respectively ({\it cf.} Fig.~\ref{fig:ResultDDNT}). We also show in the Figure, for comparison, the expected reach of the LZ experiment assuming the Standard Halo Model.  }
\label{fig:LZ-prospects} 
\end{figure}

In Fig.~\ref{fig:LZ-prospects} we show the values of the SI cross section (left panels) or SD cross section (right panels) where the maximum number of events expected at LZ is smaller than 1 (white regions), as well as the regions where the minimum number of events is larger than 1 (red regions), for all possible choices of the velocity distribution, in view of the current null search results from the direct detection experiments SuperCDMS or PICO-60 (upper panels), from neutrino telescopes (middle panels), or from considering both search strategies simultaneously (lower panels). The regions in white will remain untested, while the regions in red will be fully tested at LZ, regardless of the true velocity distribution. The regions in pink, on the other hand, produce a number of signal events larger than 1, but only for certain choices of the velocity distribution, and hence can only be partially tested. Finally, the regions in gray in the middle and bottom plots show the parameter space excluded by current experiments following our analysis in Section.~\ref{sec:two-null-results}. We also show in the Figure, for comparison, the expected reach of the LZ experiment assuming the SHM.  
Our analysis shows that LZ will be sensitive to $\sigma_{\rm SI}\gtrsim 3\times 10^{-45}\,{\rm cm}^2$ and to  $\sigma_{\rm SD}\gtrsim 3\times 10^{-40}\,{\rm cm}^2$ for $m_{\rm DM}=1\,{\rm TeV}$, regardless of the velocity distribution of dark matter particles in the Solar System, for a local density ${\rho_{\rm loc}=0.3}\,{\rm GeV}/{\rm cm}^3$. Furthermore, it shows that cross sections below $\sigma_{\rm SI}\lesssim 2\times 10^{-47}\,{\rm cm}^2$ and  $\sigma_{\rm SD}\lesssim 2\times 10^{-40}\,{\rm cm}^2$ will escape detection, even for the most favorable velocity distribution, if the dark matter mass is in the range $5-10^4\,{\rm GeV}$.

It is possible to understand analytically the contours shown in the Figure. The region where $\max\{R^{\rm (LZ)}\}(\sigma,m_{\rm DM})\leq 1$ (shown in white in the Figure) corresponds to very low cross-sections and hence none of the current upper limit constraints is saturated. As a result, the optimized velocity distribution giving the maximal recoil rate at LZ consists of a single dark matter stream $f(v)=\delta(v-v_0)$, where $v_0$ is determined from $\partial R_{v}/\partial v|_{v=v_0}=0$, if $v_0<v_{\rm max}$, and $v_0=v_{\rm max}$ otherwise, giving $\max\{R^{\rm (LZ)}\}=R^{\rm (LZ)}_{v_0}$. The white region is then defined by the condition $R^{\rm (LZ)}_{v_0}(\sigma,m_{\rm DM})\leq 1$. The red region, where $\min\{R^{\rm (LZ)}\}\geq 1$, does not exists when imposing in the minimization only the constraints from SuperCDMS: as both SuperCDMS and LZ have a non-zero velocity threshold, it is clear that the minimal rate at LZ compatible with the null search results from SuperCDMS will be equal to zero, corresponding to a single stream at zero velocity. In contrast, neutrino telescopes are sensitive to streams with velocities arbitrarily small, and therefore one expects an impact of the upper limit on the capture rate from the neutrino telescopes on the minimal rate expected at LZ. The minimal rate occurs for the velocity configurations where the upper limit on the capture rate is saturated, and therefore corresponds to two streams, which can be calculated following the lines of Subsection \ref{sec:two-null-results}, under the assumption that LZ will observe less than 1 event. Finally, the gray region in the middle plots is excluded in a halo independent manner by the null searches at neutrino telescopes only, while in the bottom plots, by the combination of the null searches at neutrino telescopes and at PandaX (left panel) or PICO-60 (right panel), as discussed in Subsection \ref{sec:two-null-results}.

\FloatBarrier

\section{Conclusions and outlook}
\label{sec:conclusions}

The interpretation of any experiment probing the dark matter distribution inside the Solar System (either through the rate of nuclear recoils, the annual modulation signal, or the neutrino flux from the Sun from dark matter particles captured in the solar interior via scatterings) is subject to our ignorance of the local dark matter density and velocity distribution. It is then important to develop methods to extract from the experimental data information about the dark matter properties, without relying on assumptions about the unknown astrophysics.  
	
In this paper we have developed a new method to calculate the minimum/maximum number of signal events in an experiment probing the dark matter distribution inside the Solar System, in view of a number of constraints from direct detection experiments and/or neutrino telescopes. The method is based in a decomposition of the velocity distribution into a linear combination of an arbitrarily large number of dark matter streams. Then, using the fact that the rate of signal events in a direct detection experiment or in a neutrino telescope is linear in the velocity distribution, we have applied methods of linear programming to optimize the rate in one experiment given a number of constraints from other experiments, and imposing that the velocity distribution must be normalized to unity. For $p$ upper limit constraints, we show that the optimized velocity distribution is composed of $p+1-r$ streams, where $r$ is the number of upper limit constraints which are not saturated. The velocities of the streams can be calculated numerically using linear programming methods, although in some simple cases the optimized values can also be derived analytically or semi-analytically.

We have illustrated our method with three concrete applications. First, we have derived a halo-independent upper limit on the spin-independent and spin-dependent cross sections from combining the null search results from the neutrino telescopes IceCube and Super-Kamiokande, assuming annihilations into $W^+W^-$ ($\tau^+\tau^-$ for $m_{\rm DM}< M_W$), and from the direct detection experiments PandaX and PICO-60, respectively. The limits we obtain are remarkably strong and reach, for ${\rm m}_{\rm DM} = 1\,{\rm TeV}$,  $\sigma_{\text{SI}}^{\text{p}}\lesssim 3\times 10^{-44}\,\text{cm}^{2}$ for the SI and $\sigma_{\text{SD}}^{\text{p}}\lesssim 10^{-39}\text{cm}^{2}$ for the SD scattering cross section, assuming $\rho_{\rm loc}=0.3\,{\rm GeV}/{\rm cm}^3$. 

Second, we have confronted the dark matter interpretation of the DAMA annual modulation signal, assuming spin-independent scattering only or spin-dependent scattering only, to the null search results from neutrino telescopes and from PandaX and PICO-60, respectively. We have found velocity distributions where the modulation signal reported by DAMA in the [2.0, 2.5], [2.5, 3.0] and [3.0, 3.5] keV energy bins are 
 compatible with the null search experiments, provided $m_{\rm DM}\gtrsim 165\,{\rm TeV}$ for the spin-independent interaction and $m_{\rm DM}\gtrsim 4.5 \,{\rm TeV}$ for the spin-dependent interaction. These solutions only arise when channeling effects in the NaI crystal are included in the analysis of the experimental results.  Further tests to these solutions, such as the requirement of reproducing not only the difference in rates between June 1st and December 1st, but also the time dependence of the modulation signal, may rule out some of them.  Moreover, these solutions are very fine tuned, and require the streams to be oriented in very concrete directions and with very little dispersion. Even small deviations from these configurations produce a number of events in PandaX or PICO-60 in excess of observations. 

Finally, we have assessed the prospects to observe dark matter induced recoils at the projected LZ experiment in view of the current null search results from IceCube, Super-Kamiokande, SuperCDMS (for the spin-independent interaction) and PICO-60 (for the spin-dependent interaction). We find that LZ will find, assuming $m_{\rm DM}=1\,{\rm TeV}$, dark matter induced recoils if $\sigma_{\rm SI}\gtrsim 3\times 10^{-45}\,{\rm cm}^2$ or  $\sigma_{\rm SD}\gtrsim 3\times 10^{-40}\,{\rm cm}^2$, regardless of the velocity distribution of dark matter particles in the Solar System (assuming $\rho_{\rm loc}=0.3\, {\rm GeV}/{\rm cm}^3$). On the other hand, values for the cross sections $\sigma_{\rm SI}\lesssim 2\times 10^{-47}\,{\rm cm}^2$ and  $\sigma_{\rm SD}\lesssim 2\times 10^{-40}\,{\rm cm}^2$ will escape detection, even for the most favorable velocity distribution, if the dark matter mass is in the range $5-10^4\,{\rm GeV}$.

This method can be extended to include in the analysis other dark matter interactions, or can be generalized to account for more realistic velocity configurations, {\it e.g.} including a smooth halo component in addition to the dark matter streams. The results of these analyses will be presented elsewhere~\cite{in-prep}.

\section*{Acknowledgments}
This work has been partially supported by the DFG cluster of excellence EXC 153 ``Origin and Structure of the Universe'' and by the Collaborative Research Center SFB1258. We are grateful to Riccardo Catena, Paolo Gondolo, Bradley Kavanagh and Sebastian Wild for useful discussions.

\section*{Note Added}
After the completion of this paper, we learned about the work \cite{Gondolo:2017jro}, where it is developed a new halo independent method and is used to estimate the unmodulated DAMA signal, based on the profiling of the likelihood function over velocity distributions.

\appendix
\section{Analytical derivation of the optimized velocity distribution}
\label{app:analytical}

In this Appendix we present an alternative method to calculate the optimized scattering rate and velocity distribution, and which can be solved analytically when the number of constraints is small. Concretely, we illustrate the method calculating the velocity distribution that minimizes the scattering rate at a direct detection experiment, given an upper limit on the capture rate from a neutrino telescope, and given the normalization constraint on the velocity distribution. To this end, we first fix the time-dependent velocity of the Earth to the value giving the smallest recoil rate. This prescription also simplifies the initial problem of calculating a three-dimensional dark matter velocity distribution, as the optimized solution depends only on the modulus of the velocity. 

The minimization problem can be formulated as:
\begin{align}
  \text{minimize}&~~ R(\{a_{v_i}\})=\sum_{i=1}^n  a_{v_i}^2 R_{v_i}\;, \\
  \text{subject to}&~~\sum_{i=1}^n a^2_{v_i}=1\;,
   \label{eq:app_norm}\\
  \text{and}&~~ C(\{a_{v_i}\})=\sum_{i=1}^n a_{v_i}^2  C_{v_i} \leq C_{\rm max}\;,
  \label{eq:app_C}
\end{align}
where we have cast the decision variables as $a_{v_i}^2$ to ensure that they are non-negative.  To minimize the objective function we introduce the Lagrangian
\begin{align}
L(\{a_{v_i}\}, \{v_i\}, s, \lambda_1,\lambda_2)=
\sum_{i=1}^n  a_{v_i}^2 R_{v_i}
-\lambda_1(\sum_{i=1}^n a^2_{v_i}-1)
-\lambda_2(\sum_{i=1}^n a_{v_i}^2  C_{v_i} +s^2 - C_{\rm max})\;,
\end{align}
with $\lambda_1$ and $\lambda_2$ Lagrange multipliers, and where $s^2$ is a (non-negative) slack variable, introduced  to recast the upper inequality constraint Eq.(\ref{eq:app_C}) into an equality constraint. The minimization conditions are:
\begin{align}
\frac{\partial L}{\partial a_{v_p}}&=2 a_{v_p}[R_{v_p}-\lambda_1 -\lambda_2 C_{v_p}]=0,~~~ p=1, ..., n \;,
 \label{eq:cond1}\\
\frac{\partial L}{\partial v_p}&= a_{v_i}^2 [\frac{\partial R_{v_p}}{\partial v_p} - \lambda_2 \frac{\partial C_{v_p}}{\partial v_p}]=0,~~~ p=1, ..., n \;,
 \label{eq:cond2}\\
\frac{\partial L}{\partial s}&=2 \lambda_2 s =0\;, \label{eq:cond3}\\
\frac{\partial L}{\partial \lambda_1}&=\sum_{i=1}^n a^2_{v_i}-1=0\;,
  \label{eq:cond4}\\
\frac{\partial L}{\partial \lambda_2}&=\sum_{i=1}^n a_{v_i}^2  C_{v_i} +s^2 - C_{\rm max}=0 \;.
\label{eq:cond5}
\end{align}

Eq.~(\ref{eq:cond3}) is satisfied either when $\lambda_2=0$  or when $s=0$. Note that the latter case corresponds, following Eq.~(\ref{eq:cond5}),  to saturating the inequality constraint, $C=C_{\rm max}$, while the former, to a strict inequality $C<C_{\rm max}$. Let us discuss each case separately.

\subsection{$C<C_{\rm max}$ (or $\lambda_2=0$)}
In this case, Eq.~(\ref{eq:cond1}) reads 
\begin{align}
a_{v_p}[R_{v_p}-\lambda_1]=0,~~~ p=1, ..., n\,,
\end{align}
which implies that only one among the $n$ decision variables, that we label as $a_{v_1}$ can be non-vanishing. Furthermore, from Eq.~(\ref{eq:cond2}) one obtains that the velocity $v_1$ is determined by the condition
\begin{align}
\frac{\partial R_{v}}{\partial v}\Big|_{v=v_1}=0\;.
\end{align}
The velocity $v$ which satisfies the previous equation corresponds to a maximum of $R_{v}$. However, extrema of the function $R_v$ may also occur at the boundaries of the region where $v$ is defined. Indeed, it is clear that a minimum arises for any stream velocity below the threshold required to induce an observable recoil, in particular for $v=0$. We then conclude that when the cross section is such that $C<C_{\rm max}$ for all streams,  one possible choice of the optimized velocity distribution corresponds to a single stream with zero velocity, giving a scattering rate equal to zero:
\begin{align}
f_{\rm opt}(v)&=\delta(v-0)\;, \nonumber \\
R_{\rm min}&=0\;.
\end{align}

\subsection{$C=C_{\rm max}$ (or $s=0$)}

In this case, the upper limit inequality Eq.~(\ref{eq:app_C}) is saturated. Then, Eq.~(\ref{eq:cond1}) reads 
\begin{align}
a_{v_p}[R_{v_p}-\lambda_1 -\lambda_2 C_{v_p}]=0,~~~ p=1, ..., n\;,
\end{align}
which implies that two decision variables, $a_{v_1}$ and $a_{v_2}$ are non-vanishing, while the remaining $n-2$ decision variables vanish. The Lagrange multipliers are easily obtained from this equation, the result being:
\begin{align}
\lambda_1=\frac{C_{v_1}R_{v_2}-C_{v_2}R_{v_1}}{C_{v_1}-C_{v_2}}\;, ~~~~
\lambda_2=\frac{R_{v_1}-R_{v_2}}{C_{v_1}-C_{v_2}}\;,
\end{align}
whereas the weights of the corresponding two streams can be found from Eqs.~(\ref{eq:cond4}) and (\ref{eq:cond5}):
\begin{align}
a^2_{v_1}=\frac{C_{\rm max}-C_{v_2}}{C_{v_1}-C_{v_2}}\;, ~~~~
a^2_{v_2}=\frac{C_{v_1}-C_{\rm max}}{C_{v_1}-C_{v_2}}\;.
\end{align}
Since $C(v)$ is a monotonically decreasing function, it follows that a solution exists only when $0\leq v_1 \leq \hat v$ and $\hat v\leq v_2 \leq v_{\rm max}$, where $\hat v$ is defined by the condition $C(\hat v)=C_{\rm max}$. 

Finally, one can determine the velocities of the two streams using Eq.~(\ref{eq:cond2}):
\begin{align}
\frac{\partial R_{v_1}}{\partial v_1} = \left(\frac{R_{v_1}-R_{v_2}}{C_{v_1}-C_{v_2}}\right) \frac{\partial C_{v_1}}{\partial v_1} \;,\\
\frac{\partial R_{v_2}}{\partial v_2} = \left(\frac{R_{v_1}-R_{v_2}}{C_{v_1}-C_{v_2}}\right) \frac{\partial C_{v_2}}{\partial v_2} \;.
\end{align}
These two equations are simultaneously fulfilled if $v_1=\bar v-\epsilon$ $v_2=\bar v+\epsilon$ with $\epsilon\rightarrow 0$, which implies, following Eq.~(\ref{eq:cond5}), that $\bar v=\hat v$. Let us denote this solution as I. 

On the other hand,  and as already mentioned for the case $C<C_{\rm max}$, the extrema may not lie on the interior of the domain where the functions are defined, in this case $0\leq v_1 \leq \hat v$ and $\hat v\leq v_2\leq v_{\rm max}$, but they may also lie at the boundary. We then find four more possible solutions, that we denote as II, III and IV and V:
\begin{align}
&{\rm solution~II:}~~v_2=v_{\rm max}, v_1~{\rm defined~by}~\frac{d}{dv}\left[\frac{C_{\rm max}}{C_{v}} R_{v}+\left(1-\frac{C_{\rm max}}{C_{v}}\right) R_{v_{\rm max}}\right]\Big|_{v=v_1}=0 \;,\\
&{\rm solution~III:} ~~v_1=0,~ v_2~{\rm defined~ by}~ \frac{d}{dv}\left[\left(\frac{C_0-C_{\rm max}}{C_0-C_{v}}\right) R_{v}\right]\Big|_{v=v_2}=0 \label{eq:solutionIII}\;, \\
&{\rm solution~IV:}~~v_1=\hat v\;, \\
&{\rm solution~V:}~~v_2=\hat v\;.
\end{align}
We note that the function $\left(\frac{C_0-C_{\rm max}}{C_0-C_{v}}\right) R_{v}$ entering in Eq.~(\ref{eq:solutionIII}) for solution III is the product of a monotonically decreasing function times a function that contains a maximum. Therefore, the minimum can only exist at the boundaries, either at $v_2=\hat v$ or at  $v_2=v_{\rm max}$. 

To summarize, the minimum of the scattering rate occurs for one of the six following velocity distributions:
\vspace{0.2cm}
\begin{itemize}
	\setlength{\itemindent}{.6in}
\item[I.] $\displaystyle{f_{\rm I}=\frac{1}{2}\delta\Big(v-(\hat v-\epsilon)\Big)+\frac{1}{2}\delta \Big(v-(\hat v+\epsilon)\Big)}$, with $\epsilon\rightarrow 0$ and $\hat v$ defined by 
\begin{align}
C_{\hat v}=C_{\rm max}\;, \nonumber
\end{align}
~~~~~~~~~~~giving a rate
\begin{align}
R_{\rm I}=R_{\hat v}\;,
\end{align}
\item[II.] $\displaystyle{f_{\rm II}=\frac{C_{\rm max}}{C_{v_1}} \delta(v-v_1)+\left(1-\frac{C_{\rm max}}{C_{v_1}}\right)  \delta(v-v_{\rm max})}$, with $v_1$ defined by 
\begin{align}
\frac{d}{dv}\left[\frac{C_{\rm max}}{C_{v}} R_{v}+\left(1-\frac{C_{\rm max}}{C_{v}}\right) R_{v_{\rm max}}\right]\Big|_{v=v_1}=0\;, \nonumber
\end{align}
~~~~~~~~~~~giving a rate
\begin{align}
R_{\rm II}=\frac{C_{\rm max}}{C_{v_1}} R_{v_1}+\left(1-\frac{C_{\rm max}}{C_{v_1}}\right) R_{v_{\rm max}}\;,
\end{align}
\item[IIIa.] $\displaystyle{f_{\rm IIIa}=\delta(v-\hat v)}$, with $\hat v$ defined by $C_{\hat v}=C_{\rm max}$, giving a rate
\begin{align}
R_{\rm IIIa}=R_{\hat v}\;,
\end{align}
\item[IIIb.] $\displaystyle{f_{\rm IIIb}=\frac{C_{\rm max}}{C_0} \delta(v-0)+\left(1-\frac{C_{\rm max}}{C_0}\right) \delta(v-v_{\rm max})}$,  giving a rate
\begin{align}
R_{\rm IIIb}=
\left(1-\frac{C_{\rm max}}{C_0}\right)  R_{v_{\rm max}}\;,
\end{align} 
\item[IV.] $\displaystyle{f_{\rm IV}=\delta(v-\hat v)}$, with  $\hat v$ defined by $C_{\hat v}=C_{\rm max}$, giving a rate
\begin{align}
R_{\rm IV}=R_{\hat v}\;,
\end{align}
\item[V.] $\displaystyle{f_{\rm V}=\delta(v-\hat v)}$, with  $\hat v$ defined by $C_{\hat v}=C_{\rm max}$, giving a rate
\begin{align}
R_{\rm V}=R_{\hat v}\;.
\end{align} 
\end{itemize}
\vspace{0.2cm}
Namely,
\begin{align}
R_{\rm min}                  
&=\min\{R_I, R_{\rm II},  R_{\rm IIIa}, R_{\rm IIIb}, R_{\rm IV}, R_{\rm V}\}\;.
\end{align}

It is clear that solutions I, IIIa, IV and V are identical. Furthermore, taking into account that the capture rate for streams decreases monotonically with its velocity, $C_{v_1}\leq C_0$, and that for zero velocity the capture rate vanishes, $R_0=0$, one obtains
\begin{align}
R_{\rm II}= \min_{v}\left\{\frac{C_{\rm max}}{C_{v}} R_{v}+\left(1-\frac{C_{\rm max}}{C_{v}}\right) R_{v_{\rm max}}\right\}\leq\left(1-\frac{C_{\rm max}}{C_{0}}\right) R_{v_{\rm max}}=R_{\rm IIIb}
\end{align}

Hence, solutions IIa, IIb, IV and V turn out to be included in solutions I and II, and the minimal rate can be simply calculated from: 
\begin{align}
R_{\rm min}=\min\{R_I,  R_{\rm II}\}\;,
\end{align}
being the optimized velocity distribution the one corresponding to the minimal rate. 

\section{Dark matter search experiments}
\label{app:experiments}

In this Appendix, we provide details about the characteristics of the experiments which are relevant to reproduce our results.

\paragraph{DAMA}
The DAMA experiment \cite{Bernabei:2013xsa,Bernabei:2008yh} searches for dark matter via the distinctive feature of an annual modulation of the event rate. The energy resolution is depicted in figure 20b of \cite{Bernabei:2008yh} and can be parametrized as follows:
\begin{equation}
\sigma(\text{E}_{\text{ee}})=(0.448 ~\text{keVee})\cdot\sqrt{\frac{\text{E}_{\text{ee}}}{\text{keVee}}}+0.0091\cdot\text{E}_{\text{ee}}\,,
\end{equation}
where $\text{E}_{\text{ee}}$ denotes the recoil energy in keVee (electron equivalent energy). Besides, we take for the efficiency in the energy bin $[E_-, E_+]$, $\epsilon_i(E_R) = \Phi \left( Q_i E_R, E_{-}, E_{+}\right)$, where $Q_i$ is the quenching factor for the isotope $i$, and $\Phi (Q_i E_R,E_{-}, E_{+})$ is the probability that an event with a nuclear recoil energy $E_R$, and hence with a quenched energy of $Q_i E_R$, is detected in the energy bin $[E_{-}, E_{+}]$. Following \cite{Savage:2008er}, we assume this probability to be Gaussian. Finally, we adopt  as quenching factors $\text{Q}_{\text{Na}}=0.30$ and $\text{Q}_{\text{I}}=0.09$ as used by the DAMA collaboration~\cite{Bernabei:1996vj}. In our analysis we also take into account the channeling effect, which was studied in detail in \cite{Bozorgnia:2010xy}.

\paragraph{PandaX}
We calculate the limits from PandaX Run 8 \cite{Tan:2016diz} and Run 9 \cite{Tan:2016zwf} by incorporating the detection efficiency from figure 2 of \cite{Tan:2016zwf} and assuming an energy threshold of $E_{\text{th}}=1.1~\text{keVnr}$. With a combined exposure of $3.3\times 10^{4}~\text{kg}\times\text{days}$, PandaX observed three events below the median of the NR calibration band, which leads to ${\cal N}_{\text{max}}=6.7$. Furthermore, we multiply the event rate by an additional factor of 0.5 to account for the fact that only half of the nuclear recoil band is used. 

\paragraph{PICO-60}
To obtain limits from the PICO-60 experiment, we follow \cite{Amole:2015lsj,Amole:2017dex} and incorporate the bubble efficiency which is given by the dashed lines in figure 4 of \cite{Amole:2015lsj}. Furthermore, we use an exposure of 1167 kg$\times$days as reported in~\cite{Amole:2017dex}. Since PICO-60 observed no signal events, we conservatively assume a 90\% C.L. upper limit on the number of recoils of ${\cal N}_{\text{max}}=2.3$.

\paragraph{SuperCDMS}
The SuperCDMS experiment \cite{Agnese:2014aze} recorded data between October 2012 and June 2013, resulting in a total exposure of 577 kg$\times$days. The energy dependent efficiency is given in figure 1 of \cite{Agnese:2014aze} and the dark matter search window is defined between 1.6 keVnr and 10 keVnr. After unblinding, SuperCDMS observed 11 candidate events which leads to an upper limit of ${\cal N}_{\text{max}}=16.6$ at 90\% confidence level.

\paragraph{LUX-ZEPLIN}
LUX-ZEPLIN (LZ) \cite{Akerib:2015cja} is the planned successor of LUX~\cite{Akerib:2012ys} and ZEPLIN~\cite{Akimov:2006qw}. LZ further refines their successful detection technique and is expected to increase the fiducial mass from 145 kg to 5.6 tonnes. After running for 1000 days, this will result in an unprecedented exposure of 5600 tonnes$\times$day. Due to the similar detector technique as LUX and Panda X, we adopt the same efficiency as in LUX, taken from figure 2 of \cite{Akerib:2016vxi} and a nuclear recoil acceptance of 0.5. 

\paragraph{Neutrino telescopes}
In order to calculate the capture rate of dark matter particles inside the sun, we use the density profile from the solar model AGSS09 \cite{Serenelli:2009yc}. For the SI interactions, we include the 29 most abundant elements inside the sun and assume the Helm form factor \cite{Lewin:1995rx}. We take into account the scattering off hydrogen and $^{14}{\rm N}$ when calculating the capture rate induced by SD interactions and we use the SD form factors provided in \cite{Catena:2015uha}. For this study, we adopt the latest results of IceCube \cite{Aartsen:2016zhm} and Super-Kamiokande \cite{Choi:2015ara}. For Super-Kamiokande, we use \texttt{DarkSUSY} \cite{Gondolo:2004sc} to convert limits on the neutrino flux induced by dark matter annihilations into limits on the capture rate.

\bibliographystyle{JHEP}
\bibliography{references}

\end{document}